\documentclass[twocolumn]{aastex7}
\usepackage{amssymb}
\usepackage{amsmath}

\shortauthors{Montano et al.}

\newcommand{\per}{\ensuremath{^{-1}}}
\newcommand{\persq}{\ensuremath{^{-2}}}

\begin{document}

\title{AGN STORM\,2. XII. Ground-Based Optical Photometry and Lag Measurements of Mrk\,817}

\author[0000-0001-5639-5484]{John W. Montano}
\email{montano3@uci.edu}
\affiliation{Department of Physics and Astronomy, 4129 Frederick Reines Hall, University of California, Irvine, CA, 92697-4575, USA}

\author[0000-0002-3026-0562]{Aaron J. Barth}
\email{barth@uci.edu}
\affiliation{Department of Physics and Astronomy, 4129 Frederick Reines Hall, University of California, Irvine, CA, 92697-4575, USA}

\author[0000-0003-1728-0304]{Keith Horne}
\affiliation{SUPA School of Physics and Astronomy, North Haugh, St.~Andrews, KY16~9SS, Scotland, UK}
\email{kdh1@st-andrews.ac.uk}

\author[0000-0002-8294-9281]{Edward M.\ Cackett}
\email{ecackett@wayne.edu}
\affiliation{Department of Physics and Astronomy, Wayne State University, 666 W.\ Hancock St, Detroit, MI, 48201, USA}

\author[0000-0003-3242-7052]{Gisella De~Rosa}
\affiliation{Space Telescope Science Institute, 3700 San Martin Drive, Baltimore, MD 21218, USA}
\email{gderosa@stsci.edu}

\author[0000-0002-0957-7151]{Y. Homayouni}
\email{ybh5251@psu.edu}
\affiliation{University of Connecticut, Department of Physics, 2152 Hillside Road, Unit 3046, Storrs, CT 06269-3046, USA}
\affiliation{Space Telescope Science Institute, 3700 San Martin Drive, Baltimore, MD 21218, USA}

\author[0000-0003-0172-0854]{Erin A.\ Kara}
\affiliation{MIT Kavli Institute for Astrophysics and Space Research, Massachusetts Institute of Technology, Cambridge, MA 02139, USA}
\email{ekara@mit.edu}

\author[0000-0002-2180-8266]{Gerard A.\ Kriss}
\affiliation{Space Telescope Science Institute, 3700 San Martin Drive, Baltimore, MD 21218, USA}
\email{gak@stsci.edu}

\author[0000-0001-8391-6900]{Hermine Landt}
\affiliation{Centre for Extragalactic Astronomy, Department of Physics, Durham University, South Road, Durham DH1 3LE, UK}
\email{hermine.landt@durham.ac.uk}
\author[0009-0000-0296-6704]{Gilvan G. Apolonio}
\email{gilvan@apolonio.com.br}
\affiliation{Department of Physics and Astronomy, N283 ESC, Brigham Young University, Provo, UT 84602, USA}

\author[0000-0003-2991-4618]{Nahum Arav}
\affiliation{Department of Physics, Virginia Tech, Blacksburg, VA 24061, USA}
\email{arav@vt.edu}
\author[0000-0001-6301-570X]{Benjamin D. Boizelle}
\email{boizellb@byu.edu}
\affiliation{Department of Physics and Astronomy, N284 ESC, Brigham Young University, Provo, UT, 84602, USA}

\author[0000-0001-9931-8681]{Elena Dalla Bont\`{a}}
\affiliation{Dipartimento di Fisica e Astronomia ``G.\  Galilei,'' Universit\`{a} di Padova, Vicolo dell'Osservatorio 3, I-35122 Padova, Italy}
\affiliation{INAF-Osservatorio Astronomico di Padova, Vicolo dell'Osservatorio 5 I-35122, Padova, Italy}
\affiliation{Jeremiah Horrocks Institute, University of Central Lancashire, Preston, PR1 2HE, UK}
\email{elena.dallabonta@unipd.it}
\author[0000-0002-4830-7787]{Doron Chelouche}
\email{doron@sci.haifa.ac.il}
\affiliation{Haifa Center for Theoretical Physics and Astrophysics (HCTPA), University of Haifa, Haifa 3498838, Israel}
\affiliation{Department of Physics, Faculty of Natural Sciences, University of Haifa, Haifa 3498838, Israel}
\author[0000-0002-0964-7500]{Maryam Dehghanian}
\affiliation{Department of Physics and Astronomy, The University of Kentucky, Lexington, KY 40506, USA}
\email{m.dehghanian@uky.edu}
\author[0000-0001-8598-1482]{Rick Edelson}
\affiliation{Eureka Scientific, Inc., 2452 Delmer Street, Suite 100,
Oakland, CA 94602, USA}
\email{rickedelson@gmail.com}
\author[0000-0003-4503-6333]{Gary J. Ferland} 
\email{gary@g.uky.edu}
\affiliation{Department of Physics and Astronomy, University of Kentucky, Lexington, KY 40506}

\author[0000-0002-2306-9372]{Carina Fian}
\email{carina.fian@inaf.it}
\affiliation{INAF – Osservatorio Astronomico di Trieste, via G.B. Tiepolo, 11, I-34143 Trieste, Italy}

\author[0000-0001-9704-690X]{Catalina Sobrino Figaredo}
\email{csobrinofigaredo@gmail.com}
\affiliation{Haifa Center for Theoretical Physics and Astrophysics (HCTPA), University of Haifa, Haifa 3498838, Israel}
\affiliation{Department of Physics, Faculty of Natural Sciences, University of Haifa, Haifa 3498838, Israel}
\author[0000-0002-2908-7360]{Michael R.\ Goad}
\affiliation{School of Physics and Astronomy, University of Leicester, University Road, Leicester, LE1 7RH, UK}
\email{mg159@leicester.ac.uk}

\author[0000-0002-9280-1184]{Diego H. Gonzalez-Buitrago}
\email{dgonzalez@astro.unam.mx}
\affiliation{Universidad Nacional Aut\'onoma de M\'exico, Instituto de Astronom\'ia, AP 106,  Ensenada 22860, BC, M\'exico}

\author[0000-0001-9457-0589]{Wei-Jian Guo}
\email{guowj@bao.ac.cn}
\affiliation{Key Laboratory of Optical Astronomy, National Astronomical Observatories, Chinese Academy of Sciences, Beijing 100012, China}

\author[]{Chen Hu}
\email{huc@ihep.ac.cn}
\affiliation{State Key Laboratory of Particle Astrophysics, Institute of High Energy Physics, Chinese Academy of Sciences, 19B Yuquan Road, Beijing 100049, People’s Republic of China}
\author[0000-0002-1134-4015]{Dragana Ili\'c}
\email{dragana.ilic@matf.bg.ac.rs}
\affiliation{University of Belgrade - Faculty of Mathematics, Department of astronomy, Studentski trg 16, 11000 Belgrade, Serbia}
\affiliation{Hamburger Sternwarte, Universit\"at Hamburg, Gojenbergsweg 112, 21029 Hamburg, Germany}
\author[0000-0003-0634-8449]{Michael D.\ Joner}
\email{joner@byu.edu}
\affiliation{Department of Physics and Astronomy, N283 ESC, Brigham Young University, Provo, UT 84602, USA}
\author[0000-0002-9925-534X]{Shai Kaspi}
\email{shaik@tauex.tau.ac.il}
\affiliation{School of Physics and Astronomy and Wise Observatory, Tel Aviv University, Tel Aviv 6997801, Israel}

\author[0000-0002-1790-3148]{Christopher~S.~Kochanek}
\affiliation{Department of Astronomy, The Ohio State University, 140 W. 18th Ave., Columbus, OH 43210, USA}
\affiliation{Center for Cosmology and Astroparticle Physics, The Ohio State University, 191 W. Woodruff Ave., Columbus, OH 43210, USA}
\email{kochanek.1@osu.edu}

\author[0000-0001-5139-1978]{Andjelka B. Kova\v cevi\'c}
\email{andjelka.kovacevic@matf.bg.ac.rs}
\affiliation{University of Belgrade - Faculty of Mathematics, Department of astronomy, Studentski trg 16, 11000 Belgrade, Serbia}

\author[0000-0002-8671-1190]{Collin Lewin}
\affiliation{MIT Kavli Institute for Astrophysics and Space Research, MIT, 77 Massachusetts Avenue, Cambridge, MA 02139, USA}
\email{clewin@mit.edu}

\author[0000-0003-3823-3419]{Sha-Sha Li}
\email{lishasha@ynao.ac.cn}
\affiliation{Yunnan Observatories, Chinese Academy of Sciences, Kunming 650216, Yunnan, People's Republic of China}

\author[0000-0001-5841-9179]{Yan-Rong Li}
\email{liyanrong@mail.ihep.ac.cn}
\affiliation{State Key Laboratory of Particle Astrophysics, Institute of High Energy Physics, Chinese Academy of Sciences, 19B Yuquan Road, Beijing 100049, People’s Republic of China}

\author[0000-0003-3086-7804]{Jun-Rong Liu}
\email{liujunrong@ihep.ac.cn}
\affiliation{State Key Laboratory of Particle Astrophysics, Institute of High Energy Physics, Chinese Academy of Sciences, 19B Yuquan Road, Beijing 100049, People's Republic of China}
\author[0000-0001-8475-8027]{Jake A. Miller}
\email{jamiller@tamu.edu}
\affiliation{Texas A\&M University, Department of Physics \& Astronomy, 400 Bizzell St, College Station, TX 77845, USA}
\author[0000-0001-7351-2531]{Jack M. M. Neustadt}
\affiliation{Department of Physics and Astronomy, Bloomberg Center, The Johns Hopkins University, Baltimore, MD 21218, USA}
\email{jneustadt@jhu.edu}

\author[0000-0002-6766-0260]{Hagai Netzer}
\affiliation{School of Physics and Astronomy, Tel Aviv University, Tel Aviv 69978, Israel}
\email{hagainetzer@gmail.com}
\author[0000-0001-5578-8614]{Paolo Ochner}
\affiliation{Dipartimento di Fisica e Astronomia ``G.\  Galilei,'' Universit\`{a} di Padova, Vicolo dell'Osservatorio 3, I-35122 Padova, Italy}
\affiliation{INAF-Osservatorio Astronomico di Padova, Vicolo dell'Osservatorio 5 I-35122, Padova, Italy}
\email{paolo.ochner@unipd.it}
\author[0000-0003-1183-1574]{Ethan R. Partington}
\affiliation{Western Kentucky University, Department of Physics and Astronomy, 1906 College Heights Blvd., Bowling Green, KY 42101, US
}
\email{epartington@ia.forth.gr}

\author[0000-0001-9585-417X]{Alessandro Pizzella}
\affiliation{Dipartimento di Fisica e Astronomia ``G.\  Galilei,'' Universit\`{a} di Padova, Vicolo dell'Osservatorio 3, I-35122 Padova, Italy}
\affiliation{INAF-Osservatorio Astronomico di Padova, Vicolo dell'Osservatorio 5 I-35122, Padova, Italy}
\email{alessandro.pizzella@unipd.it}

\author[0000-0002-2509-3878]{Rachel Plesha}
\email{rplesha@stsci.edu}
\affiliation{Space Telescope Science Institute, 3700 San Martin Drive, Baltimore, MD 21218, USA}

\author[0000-0003-2398-7664]{Luka \v C. Popovi\'c}
\email{lpopovic@aob.rs}
\affiliation{Astronomical Observatory Belgrade, Volgina 7, 11000 Belgrade, Serbia}
\affiliation{University of Belgrade - Faculty of Mathematics, Department of astronomy, Studentski trg 16, 11000 Belgrade, Serbia}


\author[0000-0002-9238-9521]{David Sanmartim}
\email{dsanmartim@lsst.org}
\affiliation{NSF NOIRLab/NSF–DOE Vera C. Rubin Observatory HQ, 950 N. Cherry Ave., Tucson, AZ 85719, USA (AURA Staff)}

\author[0000-0002-6733-5556]{Juan V.\ Hern\'{a}ndez Santisteban}
\affiliation{SUPA School of Physics and Astronomy, North Haugh, St.~Andrews, KY16~9SS, Scotland, UK}
\email{jvhs1@st-andrews.ac.uk}


\author[0000-0001-9191-9837]{Marianne Vestergaard}
\affiliation{DARK, Niels Bohr Institute, The University of Copenhagen, Jagtvej 155, DK-2200 Copenhagen N, Denmark}
\affiliation{Steward Observatory and Department of Astronomy, University of Arizona, 933 N Cherry Avenue, Tucson, AZ, 85721, USA}
\email{mvester@nbi.ku.dk}

\author[0009-0001-6008-0108]{Jack H. F. Wooley}
\email{Jack.Wooley@byu.edu}
\affiliation{Department of Physics and Astronomy, N283 ESC, Brigham Young University, Provo, UT 84602, USA}

\author[]{Sen Yang}
\email{yangsen@aynu.edu.cn}
\affiliation{College of Physics and Electrical Engineering, Anyang Normal University, Anyang, Henan 455000, China}

\author[0009-0000-1228-2373]{Zhu-Heng Yao}
\email{zhyao@bao.ac.cn}
\affiliation{National Astronomical Observatories, Chinese Academy of Sciences, 20A Datun Road, Chaoyang District, Beijing 100101, People's Republic of China}
 
\author[orcid=0000-0003-0931-0868]{Fatima Zaidouni}
\affiliation{MIT Kavli Institute for Astrophysics and Space Research, Massachusetts Institute of Technology, Cambridge, MA 02139, USA}
\email{fzaid@mit.edu}

\begin{abstract}
We present the ground-based imaging campaign and light curves of Markarian 817 as part of the multiwavelength monitoring program AGN STORM\,2. Observations were carried out over 1.4 years in \emph{uBgVriz} filters, with a median cadence of 0.4 days in \emph{g}. Reverberation lags are measured using three methods (ICCF, JAVELIN, and PyROA) with the Swift UVW2 band (1928 \AA) as the reference light curve. The ICCF centroid lags range from $3.0\pm0.8$ days for the $u$ band up to $7.9\pm1.5$ days for $z$, and are consistent with a $\tau\propto \lambda^{4/3}$ dependence, the relation expected for lamp-post reprocessing by a Shakura-Sunyaev disk. Lags measured with the other methods are systematically shorter, and deviate from a $\lambda^{4/3}$ power-law spectrum at long wavelengths.  The lags exceed thin-disk reprocessing predictions by factors of $\sim$3-6, similar to the ``disk size discrepancy'' seen in other Seyfert galaxies. We divide the campaign into three epochs with different levels of mean luminosity and X-ray obscuring column density and find that the lags vary by as much as a factor of 2 between epochs. The intrinsic spectral energy distribution is bluer and brighter during the first third of the campaign, and the longest continuum reverberation lags are obtained during that period. These results suggest that changes in ionizing luminosity can produce large variations in continuum lags on short timescales by altering the diffuse continuum luminosity emitted by the broad-line region and/or obscuring outflow, although changes in obscuration between the central engine and broad-line region may also contribute to the lag variations. 
\end{abstract}

\section{Introduction} \label{sec:intro}

Although supermassive black holes account for only $\sim$0.1\% of the bulge mass of their host galaxies, they are understood to play an outsize role in galaxy evolution through feedback effects that occur during episodes of accretion \citep[e.g.,][]{2013ARA&A..51..511K,2014ARA&A..52..589H}.  Understanding the physical processes involved in gaseous accretion onto black holes in active galactic nuclei (AGN) and the launching of winds and outflows is essential to model their impact on star formation rates and the properties and distribution of gas in the host galaxy environment. Due to the small sizes of AGN accretion flows, the most widely applicable method for studying their structure is reverberation mapping, which relies on temporal rather than angular resolution \citep{blandford1982,Peterson_1993, CACKETT2021102557}. This method uses frequent observations of AGN spanning long observing windows to resolve time delays between the light curves of continuum radiation generated close to the black hole and reprocessed line or continuum emission originating from more distant locations in the accretion disk, the broad-line region (BLR), or the obscuring torus. Reverberation mapping of broad emission lines has been essential for determination of BLR sizes, measuring black hole masses in AGN, and setting the foundation for methods used to estimate black hole masses in high-redshift quasars to trace the early growth of supermassive black holes \citep[e.g.,][]{Wandel1999,Peterson2004,Onken2004,Vestergaard2006}.

Over the past decade, ultraviolet (UV) and optical continuum reverberation mapping has become widely employed as an important probe of accretion disk structure and reprocessing physics, primarily thanks to large allocations of time for AGN monitoring programs with the Neil Gehrels Swift Observatory and with the existence of ground-based robotic telescopes including those of Las Cumbres Observatory. Early conceptions of the origin of optical continuum reverberation lags in AGN \citep[e.g.,][]{Collier1999, Sergeev2005, Cackett2007} were based on a model in which X-ray photons originating close to the black hole are absorbed by the accretion disk surface, and the absorbed energy is re-radiated with a thermal spectrum corresponding to the local disk temperature. The light-travel time between the central source and the disk reprocessing regions then gives rise to the observed reverberation time delays ($\tau$), expected to show a $\tau(\lambda) \propto \lambda^{4/3}$ delay spectrum for a \citet{1973A&A.SS} disk having a  temperature profile $T(R) \propto R^{-3/4}$. This model motivated efforts to use optical continuum reverberation mapping to measure accretion disk radii and even to use disk lags to define a distance indicator as a way to determine the Hubble constant \citep{Collier1999}.

However, data from intensive multiwavelength continuum reverberation mapping campaigns have demonstrated clear inconsistencies with the basic lamp-post reprocessing model. In the six-month AGN STORM campaign on NGC 5548 carried out in 2014, observations from the Hubble Space Telescope (HST), Swift, and ground-based facilities revealed a UV/optical lag spectrum consistent with the predicted $\tau(\lambda) \propto \lambda^{4/3}$ dependence, but with a normalization approximately three times larger than expected for disk reprocessing given the assumed BH mass and accretion rate \citep{Edelson_2015,Fausnaugh_2016}. A similar ``disk size discrepancy'' has subsequently been seen in several other objects that have been monitored by Swift and ground-based telescopes \citep[e.g.,][]{Edelson2019,Kara_2023,Prince2025,gonzalezbuitrago2025} as well as for quasars monitored through long-duration survey programs \citep[e.g.,][]{Jiang2017,Mudd2018,Jha2022,Guo_2022,Homayouni2022,2022ApJ...929...19G}, implying that either accretion disks are typically a few times larger than predicted by standard models, or that processes other than thin-disk reprocessing are responsible for the lags. These continuum reverberation mapping results reinforce earlier findings of unexpectedly large accretion disk sizes in gravitationally lensed quasars \citep{Morgan2010,Blackburne2011}.

This problem has motivated the development of a variety of models to explain the origin of optical variability and the longer-than-expected continuum lags, including models based on modifications to disk shape or reprocessing geometry \citep{Gardner2017,starkey_bowl}, changes to the disk temperature profile due to winds \citep{Sun2019}, changes to the emergent spectrum due to scattering in disk atmospheres \citep{Hall2018}, accounting for relativistic effects and allowing for a greater coronal height above the disk and boosting disk temperatures \citep{Kammoun_2021}, or including variability due to disk temperature fluctuations rather than reprocessing of coronal X-rays \citep[e.g.,][]{Cai2018,Sun2020}. X-ray illumination of the disk surface can also increase the disk temperature, increasing the radius at which reprocessed flux of a given wavelength is emitted \citep{Secunda2024}. 

Another challenge for the simple lamp-post reprocessing model stems from the fact that some monitoring programs have shown little or no correlation between short-timescale variations in X-ray and UV light curves \citep[e.g.,][]{Edelson2019,Morales2019}. This casts doubt on the premise that reprocessing of coronal X-ray emission is the primary driver of short-timescale UV/optical flux variations in AGN, although \citet{Panagiotou2022} find that the AGN STORM data for NGC 5548 can be fit with an X-ray reprocessing model that is able to successfully reproduce the variability power spectral densities in the X-rays and UV as well as the UV/optical lag spectrum. As an alternative scenario, \citet{hagendone2} propose a model in which disk reprocessing of the rapid variations of the coronal X-ray emission is only responsible for a very small fraction of the UV/optical variability. Instead, the UV/optical variations are caused by inward-propagating accretion rate fluctuations in the disk, and the reverberation lags are primarily due to reprocessing of extreme-UV photons by a wind rather than X-ray reprocessing by the disk.

Work by \citet{Korista2001}, \citet{lauther_goad_korista}, and \citet{koristagoad} highlighted the important contribution of nebular ``diffuse continuum'' (DC) emission to the UV/optical spectrum and to the continuum reverberation lags. The DC spectrum is composed of free-free and free-bound emission from photoionized gas in the BLR, and the larger size of the BLR compared with the accretion disk implies that the DC emission will show lags longer than those originating from the disk itself. Key evidence for the DC contribution to the continuum lags comes from detection of an enhanced or excess lag in the \emph{U} band. This feature was first seen in the AGN STORM campaign data for NGC 5548 \citep{Edelson_2015,Fausnaugh_2016} and subsequently detected in Swift and ground-based monitoring data for several other AGN \citep{Edelson2019,fairall9juan,Cackett2020,Vincentelli2022,Kara_2023}. HST STIS reverberation mapping data of NGC 4593 have provided the most detailed, spectrally resolved view of the enhanced lag in the \emph{U}-band region \citep{Cackett_2018_4593}. This excess is attributed to the strong bump in the DC emission spectrum shortward of the Balmer jump, and photoionization models computed for NGC 5548 predict that DC emission can account for $\sim20-40\%$ of the total continuum flux across the optical spectrum, with the largest contributions falling just shortward of the Balmer jump and the Paschen jump \citep{koristagoad}. The presence of DC emission offers a natural explanation for observed continuum lags being greater than expected from disk reprocessing alone, and some recent studies have suggested that DC emission is sufficient to fully account for the observed lags \citep[e.g.,][]{chelouche_direct_2019, Netzer2022}.

The AGN STORM 2 campaign targeted Mrk 817 for intensive multiwavelength monitoring over a 15-month span from late 2020 through early 2022. Like the earlier STORM campaign on NGC 5548 \citep{DeRosa2015}, the STORM 2 campaign was centered on a large HST program (Program ID GO-16196) of high-cadence UV spectroscopic observations. The campaign also included daily observations with Swift, ground-based optical photometry and spectroscopy, and additional X-ray observations with XMM-Newton, NICER, and NuSTAR. The overall program design and science goals, along with early results from the first three months of observations, are described in Paper I \citep{Kara_2021}. Although Mrk 817 was selected as the target for the STORM 2 program based on its historical lack of strong UV absorption lines, suggesting a clean line of sight to the central engine, the HST and X-ray observations during the initial portion of the campaign revealed blueshifted UV absorption lines and a highly absorbed X-ray spectrum, indicating the emergence of an ionized outflow.

Reverberation mapping of the broad emission lines and UV/optical continuum were central goals of the STORM 2 program. Paper II \citep{Homayouni_2023_storm2_II} presented lag measurements for UV emission lines (Ly$\alpha$, \ion{N}{5}, \ion{Si}{4}+\ion{O}{4}, \ion{C}{4}, and \ion{He}{2}).  Paper V \citep{homayouni_storm2_V} carried out more detailed measurements for \ion{C}{4}, showing that both the responsivity and the lag of the \ion{C}{4} line showed dramatic changes during different portions of the campaign, which were attributed to variations in the obscuration of the ionizing flux incident on the BLR.
Paper IV \citep{cackett_storm2_IV} presented continuum reverberation mapping results from the Swift component of the STORM 2 campaign. Key results included (1) a lack of correlation between the X-ray and UV light curves, (2) a lag spectrum consistent with the typical $\tau\propto\lambda^{4/3}$ dependence, and  (3) detection of a temporary divergence between the far-UV and near-UV light curve shapes for a brief period early in the campaign, which coincided with an increase in the absorbing column density ($N_\mathrm{H}$) as seen in X-ray observations \citep[Paper III;][]{Partington_2023_STORM2_III}. \added{An independent analysis of the Mrk 817 Swift continuum lags was also presented recently by \citet{Wang2025}.} 

An investigation of frequency-resolved continuum lags was presented in Paper VII \citep{lewin_storm2_VII}. Paper VII found that changes in lag during the campaign appear to be correlated with changes in $N_\mathrm{H}$, and proposed that the episodic launching of obscuring outflows could temporarily shield the BLR from the ionizing continuum, decreasing the DC luminosity and reducing the overall continuum lags. Paper X \citep{hagai_storm2_X} used photoionization models of BLR clouds to demonstrate that DC emission from the disk wind and from the BLR provides a good match to the lag spectrum across UV and optical wavelengths, implying that the accretion disk itself accounts for only a small portion of the observed lags.

In addition to the COS UV spectroscopic monitoring, HST STIS spectra were also obtained at several epochs in the campaign in order to monitor changes in the UV/optical spectral energy distribution (SED). Paper X used the HST COS and STIS data to show that the AGN's spectral shape is bluer when brighter. Furthermore, this change in spectral shape was found to be intrinsic to the AGN, since the bluer-when-brighter trend was too large to be caused solely by dilution of the AGN light by relatively redder host-galaxy starlight.


In this paper, we present results from the ground-based imaging component of the STORM 2 campaign, which employed 12 telescopes to monitor Mrk 817 at a daily to sub-daily cadence in seven filter bands for continuum reverberation mapping. The optical light curves described in this work have previously been incorporated into other components of the STORM 2 data analysis, including an investigation of accretion disk temperature fluctuations in Paper VI \citep{neustadt_storm2_VI}, the measurement of frequency-resolved continuum lags in Paper VII, the comparison with BLR photoionization modeling in Paper X, and the interpretation of near-IR continuum variations in Paper XI \citep{landt_storm2_IR}. The ground-based spectroscopic component of STORM 2 and broad emission-line reverberation mapping results will be presented by Hu et al.\ (in preparation).

The paper is organized as follows. Section \ref{sec:obs} details the data acquisition and facilities used, Section \ref{sec:datared} explains the photometric measurements and intercalibration of the light curves, Section \ref{sec:results} presents the lag measurements using three independent methods and cross-checks their consistency, and Section \ref{sec:flxflx} presents a flux-flux analysis and an application of the ``bowl'' disk reprocessing model. Discussion of the results and conclusions are presented in Sections \ref{sec:discussion}  and \ref{sec:conclusion}. Following Paper~I, we adopt a redshift of $z=0.031455$ from \citet{strauss_huchra_redshift} for Mrk 817, and a black hole mass of of $M_\mathrm{BH} =  3.85\times 10^7 M_\odot$ \citep{Bentz_2015}. Using historical data for Mrk 817, \citet{Kara_2021} estimated an Eddington ratio of $L/L_\mathrm{Edd} \sim 0.2$. More recently, in Paper X, \citet{hagai_storm2_X} derived an estimate of $L/L_\mathrm{Edd} \approx 0.1$ based on the 5100 \AA\ continuum flux (corrected for the contribution of DC emission) at the start of the STORM 2 campaign.

\begin{deluxetable*}{lccc}[t]
\tablecaption{Telescope and Observatory information\label{tab:telobs}}
\tablehead{
\colhead{Observatory} & \colhead{Location} & \colhead{Diameter (m)} & \colhead{\# of Visits} 
}
\startdata
LCOGT & &  &  \\
\ \ McDonald  (2 telescopes) & Texas, USA  & 1.0  &  129, 81 ($g$)\\
\ \ Teide  (2 telescopes) & La Palma, Spain & 1.0   & 31, 52 ($g$)\\
\ \ Haleakala & Hawaii, USA & 2.0  & 117 ($g$)\\
Wise & Negev, Israel & 0.46  & 285 ($g$)\\
Liverpool (LT) & Canary Islands, Spain & 2.0 & 70 ($g$)\\
Zowada & New Mexico, USA & 0.51 & 254 ($g$) \\
West Mountain (WMO) & Utah, USA &  0.92  & 43 ($V$)\\
Calar Alto (CAHA) & Andaluc\'ia, Spain &  2.2  & 89 ($V$)\\
Yunnan - Lijiang Station & Yunnan Province, China & 2.4  & 49 ($V$)\\
Asiago  & Asiago, Italy &  0.91  & 21 ($V$)\\
\enddata
\tablecomments{The number of visits listed for each telescope denotes the number of nights in which observations were obtained in the $g$ or $V$ bands, after the removal of bad epochs and outliers from the light curves. }
\end{deluxetable*}

\section{Data} \label{sec:obs}
\subsection{Ground-based Observations}
The ground-based campaign used the observatory facilities listed in Table \ref{tab:telobs}. Table \ref{tab:cams} summarizes the camera properties of each telescope. In total 12 telescopes were used for the ground-based campaign, of which five are part of the Las Cumbres Observatory network. Additional details for each observatory are given below. The STORM\,2 ground-based campaign began in 2020 November and ended in 2022 April. The space-based campaign concluded earlier, in 2022 February. The ground-based imaging and spectroscopic campaign was planned to extend for a longer duration in order to continue monitoring the optical continuum and broad-line response to any significant UV variations that might occur near the end of the Swift campaign. Intensive space and ground-based monitoring of Mrk 817 continued for two additional years after the conclusion of the STORM\,2 campaign as part of an extended monitoring campaign that will be presented in future work.

We obtained data using the Johnson/Bessell \emph{B} and \emph{V}, Sloan Digital Sky Survey (SDSS) \emph{u$^\prime$g$^\prime$r$^\prime$i$^\prime$z$^\prime$}, and Pan-STARRS \emph{z$_s$} filters. The SDSS $z^\prime$ and Pan-STARRS \emph{z$_s$} light curves were combined, since we found no significant difference in the light curve shape or lag between them. For brevity the SDSS and Pan-STARRS bands will be referred to as \emph{ugriz} except when distinguishing between $z^\prime$ and $z_s$. Table\,\ref{tab:cams} lists the specific filters used on each telescope.

Observations with the \textit{Las Cumbres Observatory Global Telescope} network \citep[LCOGT;][]{2013PASP..125.1031B} were obtained through the AGN key project ``Intensive Disk Reverberation Mapping of Nearby Active Galactic Nuclei'' (program LCO-2020B-003). Three of LCOGT's northern sites were used:  McDonald Observatory in Texas (two 1.0 m telescopes, V37 and V39), Teide Observatory in the Canary Islands (two 1.0 m telescopes, Z24 and Z31) and Haleakala Observatory on Maui (the 2.0 m Faulkes Telescope North; FTN F65). The McDonald site observed throughout the entire campaign. The Teide site began observing in 2021 July and continued through the remainder of the campaign, while the FTN 2.0 m observed from 2021 April until the end of the campaign. The FTN employed the Multicolor Simultaneous Camera for studying Atmospheres of Transiting exoplanets (MuSCAT3) multi-channel imager \citep{10.1117/12.2559947}, which provided higher signal-to-noise (S/N) images in the \emph{griz} bands than the other telescopes. All images were delivered via the LCOGT archive system\footnote{\url{https://archive.lco.global}} where they are processed through the BANZAI pipeline to perform bad-pixel masking, bias subtraction, dark subtraction, flat-fielding, and astrometric calibration \citep{curtis_mccully_2018_1257560}.

The \textit{Liverpool Telescope} \citep[LT;][]{10.1117/12.551456} is a robotic 2.0 m telescope that observed Mrk\,817 in the \emph{ugriz} bands between 2021 January and 2022 January. The images were processed by the standard LT reduction pipeline.

The \textit{Dan Zowada Memorial Observatory} \citep{Carr_2022} uses a 20-inch robotic PlaneWave telescope that provided the campaign with high-cadence imaging in the \emph{ugriz} filters. However, we  excluded the Zowada \emph{u}-band data from our light curves since the S/N was substantially lower than the \emph{u}-band data obtained at other sites. \added{The Zowada observations included in this work began in 2020 November and continued through the end of the campaign. These light curves have also been measured and analyzed independently by \citet{Miller2026}, including an earlier monitoring season spanning 2018-2019. }

At the \textit{Wise Observatory}, the Centurion 18-inch telescope \citep{brosch_centurion_2008} provided near-daily cadence observations in the \emph{griz} filters. However, the \emph{z}-band data was excluded due to its low S/N. Images were processed for bias and dark subtraction and flat-fielding using IRAF routines. Observing ran from 2020 November through 2022 April. 

The \textit{West Mountain Observatory} (WMO) BYU 0.9-m reflector observed Mrk\,817 in the \emph{B} and \emph{V} filters. All images from WMO were processed using IRAF for overscan, bias subtraction, dark subtraction, flat-fielding, and astrometric calibration. WMO observed Mrk\,817 from the start of the campaign until 2021 December. Additional data were obtained in the Johnson-Cousins \emph{R}-band filter, but we did not use the \emph{R}-band data due to the substantial difference in passbands between the Johnson-Cousins \emph{R} and the SDSS $r^\prime$ filters used at the other facilities.

The \textit{Calar Alto Observatory} (CAHA) 2.2 m telescope and the \textit{Yunnan Observatory} Lijiang (LJ) 2.4 m telescope observed Mrk\,817 in the \emph{V} band. Observations at these facilities ran from November 2020 through May 2022. 

Imaging with the \textit{Asiago Observatory} (AS) 0.92 m telescope in the \emph{BVugri} bands occurred from 2021 October through December. Due to the relatively high latitude of the Asiago site ($45^\circ52\arcmin$), these observations were able to fill in the seasonal gap when most other sites were not able to observe Mrk\,817. 

Exposure times ranged from 45 s to 300 s depending on telescope and filter. At most sites, two exposures were taken per filter for each nightly visit. After combining the data from all the telescopes, the median observing cadence for each filter is: 0.9 days in $u$ (379 visits), 0.9 days in $B$ (302 visits), 0.4 days in $g$ (1019 visits), 0.9 days in $V$ (468 visits), 0.4 days in $r$ (1048 visits), 0.4 days in $i$ (934 visits), and 0.7 days in $z$ (724 visits). These visit counts correspond to the number of epochs in the final light curves after rejection of outliers caused by poor weather or instrument problems.

\begin{deluxetable*}{lccccc}
\tablecaption{Camera and Observation Properties\label{tab:cams}}
\tablehead{
\colhead{Label} & \colhead{Camera--CCD} & \colhead{Pixel Scale (\arcsec/pix)} & \colhead{FOV} & \colhead{Filters} & \colhead{\added{Median Seeing}}
}
\startdata
LCOGT-V37 & Sinistro & 0.389 & $26\farcm5\times26\farcm5$ & \emph{BVu$^\prime$g$^\prime$r$^\prime$i$^\prime$z$_s$} & 1\farcs7 ($g$)\\
LCOGT-V39 & Sinistro & 0.389 & $26\farcm5\times26\farcm5$ & \emph{BVu$^\prime$g$^\prime$r$^\prime$i$^\prime$z$_s$} & 1\farcs5 ($g$)\\
LCOGT-Z31 & Sinistro & 0.389 & $26\farcm5\times26\farcm5$ & \emph{BVu$^\prime$g$^\prime$r$^\prime$i$^\prime$z$_s$} & 1\farcs4 ($g$)\\
LCOGT-Z24 & Sinistro & 0.389 & $26\farcm5\times26\farcm5$ & \emph{BVu$^\prime$g$^\prime$r$^\prime$i$^\prime$z$_s$} & 1\farcs4 ($g$)\\
LCOGT-F65 & MuSCAT3 & 0.270 & $9\farcm1\times9\farcm1$ & \emph{g$^\prime$r$^\prime$i$^\prime$z$_s$} & 1\farcs1 ($g$)\\
Wise & QSI 683 (KAF-8300) & 0.882 & $48.9^\prime\times36.8^\prime$ & \emph{g$^\prime$r$^\prime$i$^\prime$z$^\prime$} & 2\farcs3 ($g$)\\
WMO & FLI PL-3041 & 0.490 & $25\farcm2\times25\farcm2$ & \emph{BVR} & 1\farcs2 ($V$)\\
Liverpool & IO:OO &  0.304 & $10^\prime \times 10^\prime$ & \emph{u$^\prime$g$^\prime$r$^\prime$i$^\prime$z$^\prime$} & 0\farcs9 ($g$) \\
Zowada & SBIG STL-1001e & 0.882 &  $30^\prime\times30^\prime$ &\emph{u$^\prime$g$^\prime$r$^\prime$i$^\prime$z$_s$} & 3\farcs0 ($g$)\\
CAHA & 2Kx2K & 0.530 & $11^\prime\times11^\prime$ &\emph{V} & 1\farcs0 ($V$)\\
Lijiang & E2V42-90 & 0.286 & $10^\prime \times 10^\prime$ &\emph{V} & 1\farcs6 ($V$)\\
Asiago & ON-Semi KAF-16803 & 0.870 & $59^\prime\times59^\prime$ & \emph{Vu$^\prime$r$^\prime$i$^\prime$} & 3\farcs0 ($V$)\\
\enddata
\tablecomments{\added{Median seeing is listed for the \emph{g} band, or otherwise for the \emph{V} band for telescopes without \emph{g}-band observations.}}
\end{deluxetable*}

\begin{figure}
    \centering
    \includegraphics[width=\linewidth]{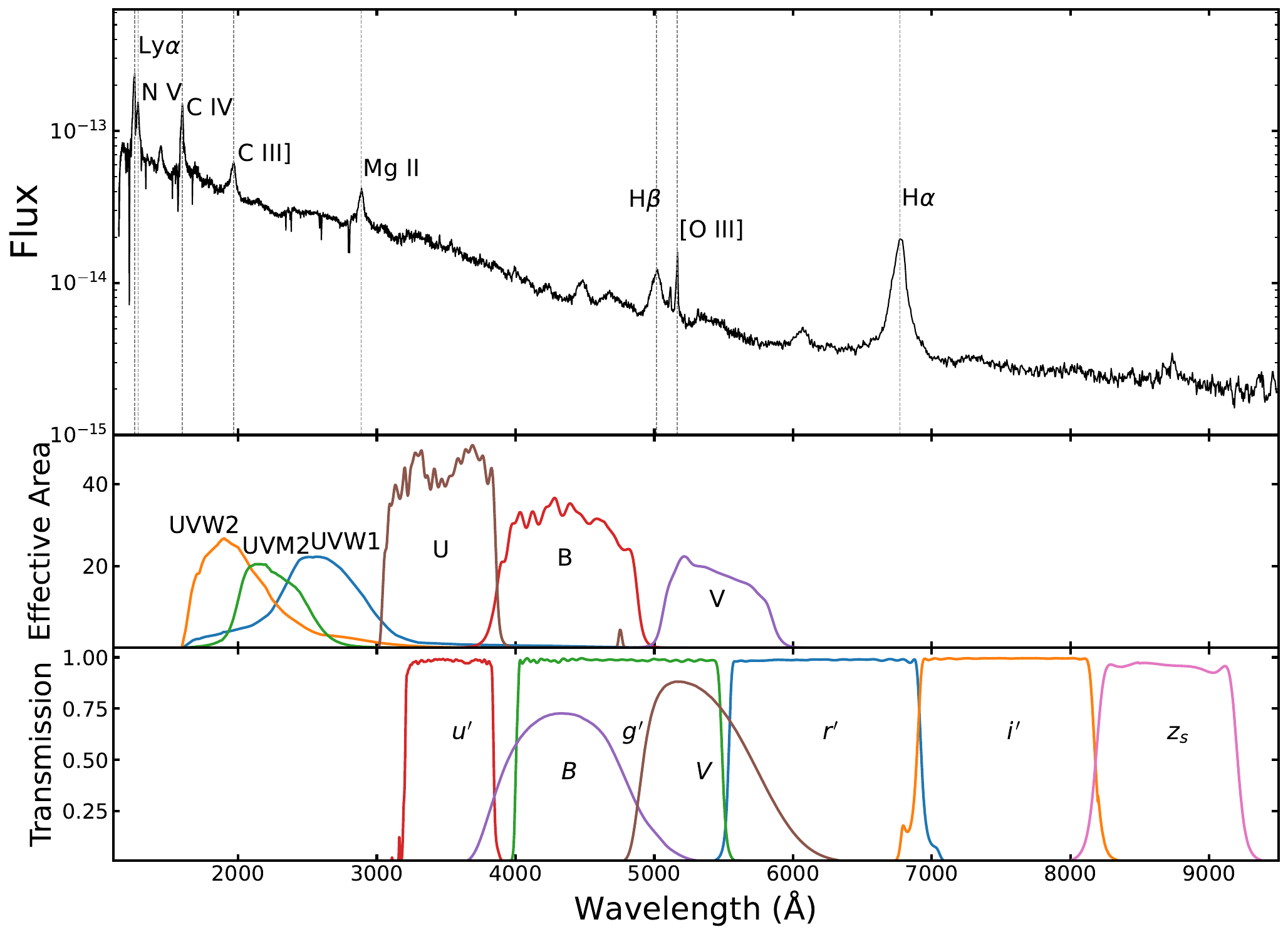}
    \caption{Top: HST STIS spectrum of Mrk\,817 observed  on 2022 January 2, in units of erg cm$^{-2}$ s$^{-1}$ \AA$^{-1}$. Middle: Effective area curves for the Swift UVOT filter bands. Bottom: Transmission curves for the LCO filter set.}
    \label{fig:specfilter}
\end{figure}

Figure\,\ref{fig:specfilter} illustrates the Swift and ground-based filter passbands\footnote{Filter transmission curves are taken from \url{https://lco.global/observatory/instruments/filters/} for LCO, and \url{http://svo2.cab.inta-csic.es/theory/fps/} for Swift.} in comparison to an HST STIS UV/optical spectrum of Mrk 817 taken on 2022 January 02 as part of the STORM 2 program (de Rosa et al., in preparation).\footnote{Based on observations made with the NASA/ESA Hubble Space Telescope, obtained at the Space Telescope Science Institute, which is operated by the Association of Universities for Research in Astronomy, Inc., under NASA contract NAS5-26555. These observations are associated with program \#16196.} The STIS spectrum is composed of observations with the G140L, G230L, G430L, and G750L gratings taken with the 0\farcs2-wide STIS slit. The strong H$\alpha$ emission line falls at the red end of the $r$ filter passband, with a small contribution to the $i$ band, while the $B$, $g$, and $V$ bands overlap the region containing H$\beta$ and prominent \ion{Fe}{2} blends. The $u$ band includes a substantial contribution from the ``small blue bump'' consisting of Balmer continuum and \ion{Fe}{2} emission. Additionally, the Swift UVW2 band, used as the reference band for lag measurements, includes the \ion{C}{3}] $\lambda$1909 \AA\ emission line. 
Future work will present detailed model fits to the HST spectra that can constrain the relative contributions of these spectral components to each of the broad-band filters.

\subsection{Additional processing steps} The facility data reduction pipelines for each instrument included basic steps such as bias subtraction and flat-fielding as described above, and in most cases also added World Coordinate System (WCS) information to the image headers. For a subset of the telescopes, we carried out additional processing steps as described below to prepare the data for photometric measurements. 

\emph{Fringe correction:} Most of the telescopes show some degree of fringing in the \emph{i} and \emph{z}-band images. This can be addressed by subtracting a fringe frame. However, we only carried out fringe correction for the Zowada \emph{z}-band data where the fringing was particularly severe. We created a fringe flat frame by combining 20 exposures of random blank sky fields and removed sources by clipping $>$$3\sigma$ outliers. A scaled version of the fringe flat was subtracted from each \emph{z}-band image of Mrk\,817, where the normalization of the fringe flat was determined by finding the scale factor that minimized the background sky residuals in the fringe-subtracted frame. This effectively removed the fringe pattern and reduced the scatter in the Zowada \emph{z}-band light curve. 

\emph{Astrometric calibration:} The instrument pipelines for Lijiang and Calar Alto did not include a determination of an astrometric solution. For images from these facilities, we used \texttt{astrometry.net} \citep{Lang_2010} to determine the WCS.

\section{Light Curve Measurements} \label{sec:datared}
Photometric reverberation mapping campaigns often include data from multiple telescopes that have different camera, detector, and filter properties resulting in inhomogeneous data sets that require extra care to produce well-calibrated light curves. We created an aperture photometry pipeline designed specifically for AGN photometric monitoring campaigns such as STORM\,2. The method was based in part on an earlier IDL-based pipeline for AGN photometry described by \citet{Pei_2014}. Our pipeline is written in Python 3 and is based on Astropy routines \citep{astropy:2013,astropy:2018,astropy:2022}. It incorporates the PyCALI method of \citet{Li_2014} to intercalibrate the light curves obtained from different telescopes. This pipeline has previously been used for the first results of the Mrk\,817 STORM\,2 campaign \citep{Kara_2021}, as well as campaigns on NGC\,4395 \citep{Montano_2022}, Mrk\,335 \citep{Kara_2023}, and PG 1302-102 \citep{Liu:2024nve}. Below we provide a detailed description of the method.

The first step is ensuring that a uniform system of time is applied to all of the image headers. We first calculate the Heliocentric Julian Date (HJD) and Modified Julian Date (MJD) for the midpoint of each exposure using the information in each image header. For the STORM\,2 campaign we use HJD as our time standard and in this paper we list dates as a truncated HJD defined as THJD = HJD $-$ 2450000. The date conversion is performed via the PyAstronomy package \citep{pya}.\footnote{\url{https://github.com/sczesla/PyAstronomy}}

For a given AGN field, the code takes as an input a list of coordinates  (right ascension and declination)  for the AGN and  a set of comparison stars in the field. The same list of comparison stars is used for all filter bands. In each image, the AGN and comparison stars are identified by their coordinates using the Astropy WCS package, and their precise pixel coordinates are then determined by \added{finding their flux-weighted centroid}.  To ensure that the centroid is determined accurately we perform two iterations of centroiding for each object, first with a 10\arcsec\ square box and then with a 5\arcsec\ square box to refine the centroid position.  \added{From simulations carried out for Gaussian sources of FWHM 1\farcs5 and S/N=100 sampled at a pixel scale of 0\farcs389 (matching that of the LCOGT Sinistro cameras), we found that the median centroiding error using this method is $\approx0\farcs07$, so the typical centroiding errors are very small compared with the 5\arcsec-radius photometric aperture.} For Mrk\,817 field we used 5 comparison stars, although only 3 or 4 of the comparison stars were used for some cameras with smaller fields of view.

\begin{figure*}
    \includegraphics[width=.99\linewidth]{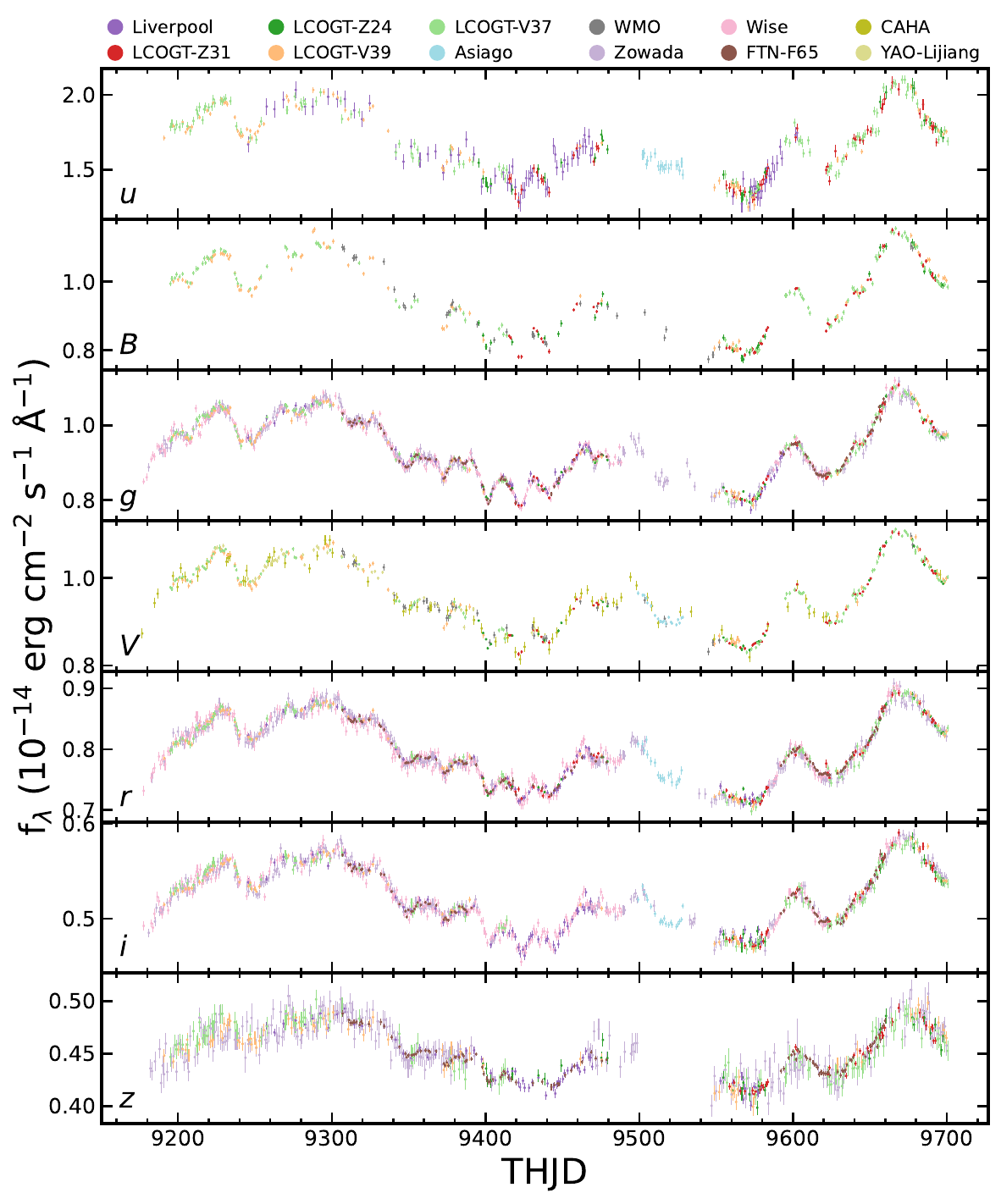}
    \caption{Mrk\,817 light curves for the STORM\,2 campaign including data from all telescopes, after averaging of multiple data points from the same visit, intercalibration, error expansion, and outlier rejection. Data points are color-coded by telescope as shown at the top.}
    \label{fig:bigplot}
\end{figure*}

For the photometric measurements, we use the \texttt{aperture\_photometry} function within the \texttt{photutils.aperture} package of AstroPy, employing the default \texttt{method='exact'} option to calculate fractional-pixel contributions at the edges of the photometric aperture.  We adopt a standard circular aperture of radius 5\arcsec\ and a background annulus spanning 15\arcsec--20\arcsec\  for sky level estimation across all image sets. To determine the sky level within the background annulus we follow a method similar to that used by \emph{DAOphot} \citep{1987PASP...99..191S} to derive an estimate of the mode sky value, taken to be three times the median pixel value minus two times the mean value, after clipping of outlier pixels. The derived sky value is then scaled by the number of pixels in the photometric aperture and subtracted from the aperture measurement. The photometric uncertainty is determined by the photon statistics and read noise within the aperture. \added{The 5\arcsec\ photometric aperture radius was chosen to be the same as that used for the STORM\,2 Swift UVOT photometry \citep{cackett_storm2_IV}. This provides consistency between the ground-based and space-based photometry in the contribution of host galaxy starlight to the photometric aperture, which is beneficial for the flux-flux analysis (Section \ref{sec:flxflx})}.

Due to transparency variations and airmass differences among the observations, the raw photometry from each night must be normalized to a consistent flux scale. For a given filter we carry out the normalization separately for each telescope and then merge the data from all of the telescopes at a later stage. The photometric measurements are normalized by determining a scale factor to apply to the measurements from each image. We calculate the scale factors by simultaneous fitting of all the scaled comparison star light curves assuming that the comparison stars are intrinsically non-variable. Using the comparison star photometry we calculate a scale factor $s$ for each image by minimizing

\begin{equation}
\chi^2 = \sum_{i=1}^K \sum_{j=1}^N \frac{(s_{j} f_{ij} - \Bar{f_i})^2}{(s_{j}\sigma_{ij})^2},    
\end{equation}
where $K$ is the number of comparison stars, $N$ is the number of images, $s_{j}$ is the scale factor for image $j$, $f_{ij}$ is the count rate and $\sigma_{ij}$ is the uncertainty for star $i$ in image $j$, and $\Bar{f_i}$ is the mean count rate for star $i$ across the time series. After determining the optimal scale factor for each image, the count rate and uncertainties for the AGN and comparison star are multiplied by that scale factor to obtain the scaled light curves. 

To convert the scaled count rates to flux units we obtain comparison star magnitudes from the APASS catalog \citep{2018AAS...23222306H}. These magnitudes are then converted to units of erg cm\persq\ s\per\ \AA\per\ using zeropoints from \citet{1996AJ....111.1748F} and \citet{1998A&A...333..231B}.
For each telescope and filter, we calculate a conversion factor from the count rate to $f_\lambda$ by determining the conversion factor individually for each comparison star based on its mean count rate in the scaled light curve. The mean conversion factor for all comparison stars (after removal of any $>5\sigma$ outliers) is then applied to the AGN light curve.

Multiple measurements from a given telescope taken within a window of $<8$ hours are considered to be a single ``visit'' and combined by weighted averaging to produce a single data point. The pipeline saves two versions of the AGN light curve, one including each individual photometric measurement and one with the nightly averaging applied. For the analysis and figures in this paper we use the nightly averaged light curves. 

A common problem in AGN photometry based on data from multiple telescopes is that differences in wavelength-dependent throughput for different telescopes, combined with the color differences between the AGN and the comparison stars, can result in slight offsets in the overall flux scales of the AGN light curves even after applying the above procedures.  These offsets can be noticeable even between telescopes having the same camera and filter designs, such as the LCOGT 1\,m network \citep[e.g.,][]{fairall9juan}. Additionally, the measurement uncertainties at this stage are based on photon-counting and detector noise statistics and do not account for sources of systematic error such as seeing variations or flat-fielding errors. To address these issues we apply an intercalibration procedure to bring the flux scales from different telescopes into agreement and adjust the error bars to account for additional systematics. 

For intercalibration we use PyCALI, a code that applies an additive shift and a multiplicative scale factor to align each telescope's light curve to a common flux scale \citep{Li_2014}. PyCALI employs a damped random walk (DRW) to model the AGN's variability along with a Markov Chain Monte Carlo (MCMC) process to find the best parameters and uncertainties.

We implemented two modifications compared with the typical application of PyCALI. The first is that we apply PyCALI to the data twice in order to obtain a better DRW model fit. We found that due to the underestimated total errors on the photometric data points, the DRW model fit from the first PyCALI run exhibited too much high-frequency structure as the model strained to fit the individual data points. After the first PyCALI run, we expanded the errors on the data points following the procedure described below and ran PyCALI again on the intercalibrated light curves. This produced smoother DRW model fits that provided a better intercalibration of the light curves. 

The second modification was to expand the errors on the data points through an additional error term added in quadrature. The additional error term is determined by finding the value that yields $\chi^2_{\textrm{dof}} = 1$ with respect to the DRW model. This additional error term was determined individually for each telescope and filter. After the first PyCALI run and error expansion we rejected data points that were $>3\sigma$ outliers with respect to the DRW model fit. For the \emph{g} band, $0.8\%$ of data points were rejected.  After the second PyCALI run, the errors were adjusted again with the same procedure to obtain $\chi^2_{\textrm{dof}} = 1$ for each telescope and filter individually. By applying the error bar expansion separately to each telescope's data, this procedure preserves the high S/N of the best-quality light curves, such as those from the FTN 2\,m telescope in the \emph{griz} bands. \added{This procedure substantially expanded the estimated uncertainties compared to the original photometric estimates. For example,  the median photometric S/N for the original  LCOGT \emph{g}-band light curve prior to intercalibration is 394, and after error bar expansion the median S/N is 
151.} In the final, error-expanded light curves, the median S/N (over all telescopes) is  52, 143, 126, 196, 110, 123, and 53, respectively, in the \emph{uBgVriz} bands.

The final intercalibrated, uncertainty-expanded light curves are shown in Figure\,\ref{fig:bigplot} and the results are presented in Table \ref{table:photometry}.

\begin{deluxetable}{lCCCC}[t!]
\tablecaption{Mrk\,817 Photometry Data Table}
\tablehead{\colhead{Band} & \colhead{HJD$-2450000$} & \colhead{$f_\lambda$} & \colhead{$\sigma(f_\lambda)$} & \colhead{Tel}\\
\colhead{} &\colhead{} & \multicolumn{2}{c}{($10^{-15}$ erg cm\persq\ s\per\ \AA\per)} &\colhead{} }
\startdata
\emph{g} & 9177.6342 & 0.8492 & 0.0079 & \textrm{Wise}\\
\emph{g} & 9179.6347 & 0.8609 & 0.0101 & \textrm{Wise}\\
\emph{g} & 9180.9856 & 0.8880 & 0.0117 & \textrm{Zowada}\\
\emph{g} & 9182.3568 & 0.9091 & 0.0073 & \textrm{Wise}\\
\emph{g} & 9183.6165 & 0.9204 & 0.0096 & \textrm{Wise}\\
\emph{g} & 9184.6137 & 0.9167 & 0.0100 & \textrm{Wise}\\
\emph{g} & 9186.6141 & 0.9442 & 0.0094 & \textrm{Wise}\\
\emph{g} & 9187.6282 & 0.9429 & 0.0086 & \textrm{Wise}\\
\emph{g} & 9188.0049 & 0.9336 & 0.0126 & \textrm{Zowada}\\
\emph{g} & 9188.6245 & 0.9312 & 0.0079 &\textrm{Wise}\\
\enddata
\label{table:photometry}
\tablecomments{ The HJD values are the midpoint of each exposure. This table is published in its entirety in  machine-readable format that includes all filters. A portion is shown here for guidance regarding its form and content.}
\end{deluxetable}
\section{Lag Analysis and Results} \label{sec:results}

\begin{figure*}
    \centering
    \includegraphics[width=\linewidth]{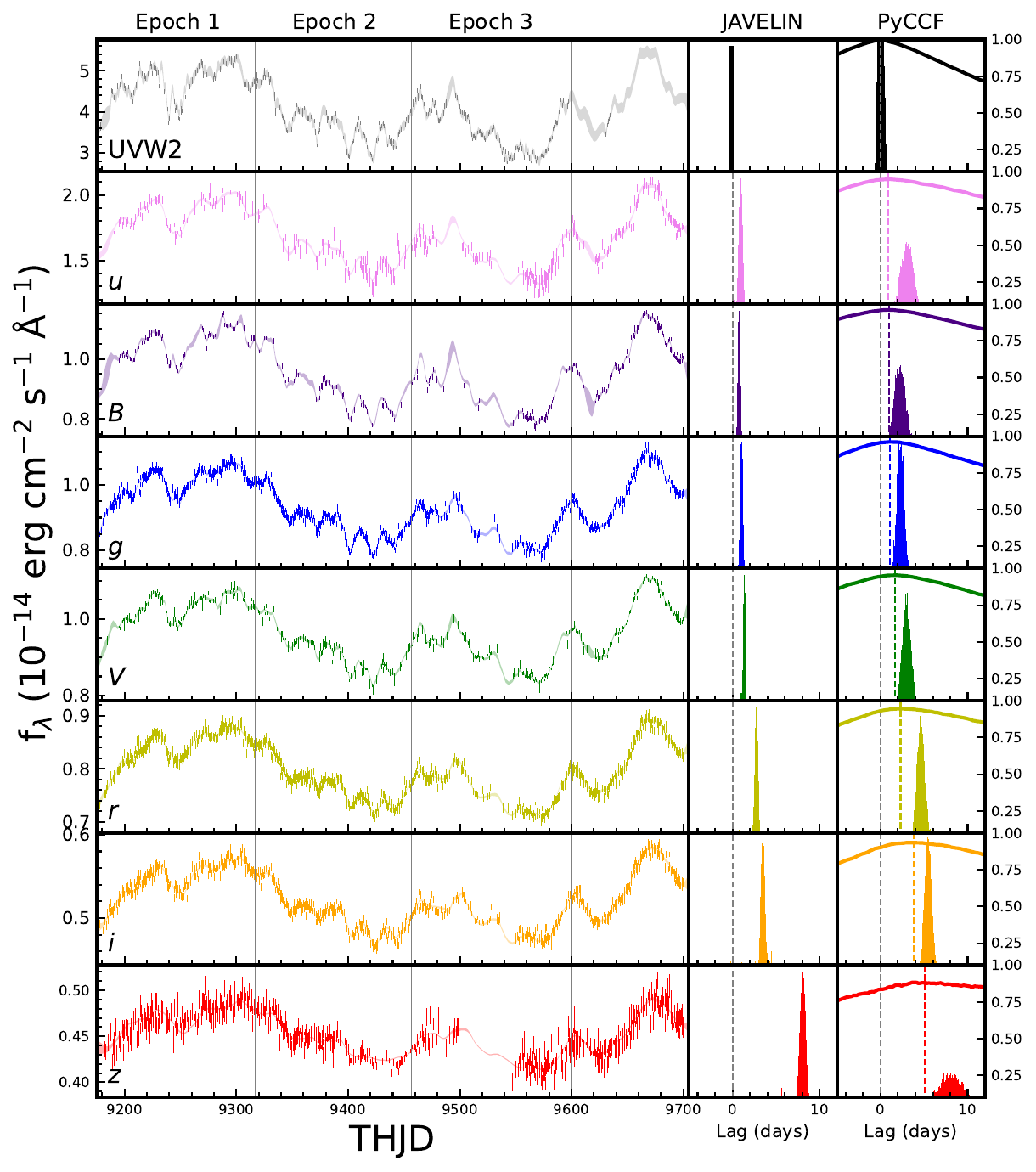}
    \caption{Left panels present the Mrk\,817 light curves and JAVELIN DRW models (shaded curves) for the ground-based filters and Swift UVW2-band that is used as the driving band for lag measurements. The Swift UVW2 DRW model is from JAVELIN fitting to the UVW2- and $u$-bands. The 2nd to the right panel shows the JAVELIN posterior distributions, the right-most panel show the cross-correlation curve of each band to the UVW2 light curve, and CCF centroid distributions. The gray vertical dashed lines denote zero lag, and colored dashed lines indicate the centroid of the CCF peak lag distribution $\tau_{\mathrm{peak}}$. The thin gray vertical lines in the light curve panels mark the boundaries between the three epochs as defined by \citet{lewin_storm2_VII}. The final segment of the light curves, after THJD 59600, was not used in the \citet{lewin_storm2_VII} analysis since only six Swift observations were obtained during that period. \added{Swift monitoring of Mrk 817 continued after THJD 9635 as part of a separate observing program (the extended campaign), and those data will be presented in future papers.}}
    \label{fig:javmodels}
\end{figure*}

\subsection{Lag Measurements} \label{sec:lags} 
We measure lags using three independent methods: ICCF \citep{Gaskell,White_1994PASP..106..879W}, JAVELIN \citep{2010ascl.soft10007Z}, and PyROA \citep{pyroa}.  Each measurement method is described below.

Our analysis of the continuum lags also incorporates the UV continuum light curves measured from the HST COS spectra from Paper II \citep{Homayouni_2023_storm2_II} and the Swift UVOT light curves from Paper IV \citep{cackett_storm2_IV}.\footnote{To distinguish between the Swift UBV and ground-based \emph{uBV} bands, we use Roman font for Swift filters and italics for ground-based filters.} For the HST and Swift UVOT lags we adopt the values from Paper IV. As a consistency check, we re-measured the lags for the HST and Swift bands with each method and found results in close agreement (within $1\sigma$) with those in Paper IV. All lags are measured with the Swift UVW2 light curve as the reference/driving light curve. We also tested measuring lags with the HST 1180\,\AA\ light curve as the reference band, but found that the Swift UVW2 band produced better results due to the higher temporal cadence of the Swift light curve. However, UVW2 is not a pure continuum band, as emission features (primarily \ion{C}{3}]) account for $\sim20-25\%$ of the total UVW2 flux based on preliminary model fits to the HST STIS spectra (de Rosa et al., in prep.). This emission-line contribution is time-lagged relative to the underlying continuum in the UVW2 light curve, and its potential impact on lag measurements has not generally been explored in prior Swift-based reverberation mapping studies.

\subsubsection{ICCF} \label{sec:pyccf}
Reverberation lags are calculated for each filter band using the standard interpolated cross-correlation (ICCF) method, using the code PyCCF \citep{peterson_pyccf,2018ascl.soft05032S}. This method uses linear interpolation across gaps between data points to calculate the correlation between unevenly sampled time series, with lag uncertainties estimated by flux randomization and random subset sampling (FR/RSS).  We use $N = 10000$ iterations with an interpolation sampling interval of 0.1 days, and a lag search range of $\pm30$ days. The adopted lag $\tau_{\textrm{cen}}$ is the median of the CCF centroid distribution with 1$\sigma$ uncertainties reported.  For each iteration, $\tau_{\textrm{cen}}$ was calculated using points in the CCF above 0.8\,$r_{\rm{max}}$ where $r_{\textrm{max}}$ is the peak amplitude of the CCF. The CCFs and CCF centroid distributions are shown in Figure\,\ref{fig:javmodels}. We also report $\tau_\mathrm{peak}$, the centroid of the distribution of CCF peak lags from the FR/RSS iterations. All the optical light curves show strong correlations with the UVW2 light curve, with peak correlation coefficients of $r_\mathrm{max}=0.94-0.96$ for the $u$ through $i$ bands, and 0.88 for the $z$ band which has lower S/N and more host galaxy contamination than the shorter-wavelength bands.

\subsubsection{JAVELIN} \label{sec:javelin} 

Just Another Vehicle for Estimating Lags In Nuclei \citep[JAVELIN;][]{Zu_2011, Zu_2013} estimates the time lag delays by modeling the light curves as a DRW and assuming that the driving and responding light curves are related by a top-hat transfer function. This process uses an MCMC method to determine posterior distributions for parameters including the lag and the DRW damping timescale. We first constrained the continuum using the UVW2 light curve with $nwalkers=1000,nchain=1000,nburn=1000$. We fit pairs of light curves rather than fitting all light curves simultaneously (always using UVW2 as the driving band) as we found that the computational time was prohibitive when fitting more than two bands at once, due to the large number of data points in the light curves.

The JAVELIN DRW light curve models are shown in Figure\,\ref{fig:javmodels} along with the posterior distributions of the lags. 

\subsubsection{PyROA}
PyROA is a Python package that uses a Running Optimal Average (ROA) algorithm to model the light curves. The free parameters of the PyROA model include the light curve mean $A_i$ for band $i$, the light curve rms $B_i$, the width for the ROA function $\Delta$, the time lag $\tau_i$, and the extra error $\sigma_i$.  The extra error term in our fits always converged to the minimum allowed value since the light curve uncertainties were already adjusted. PyROA uses an MCMC method to find the best parameters and uncertainties. We use \emph{Nsample} = 10000 and \emph{Nburnin} = 5000, with default priors except for the lag range prior.

\begin{figure*}
    \centering
    \includegraphics[width=\linewidth]{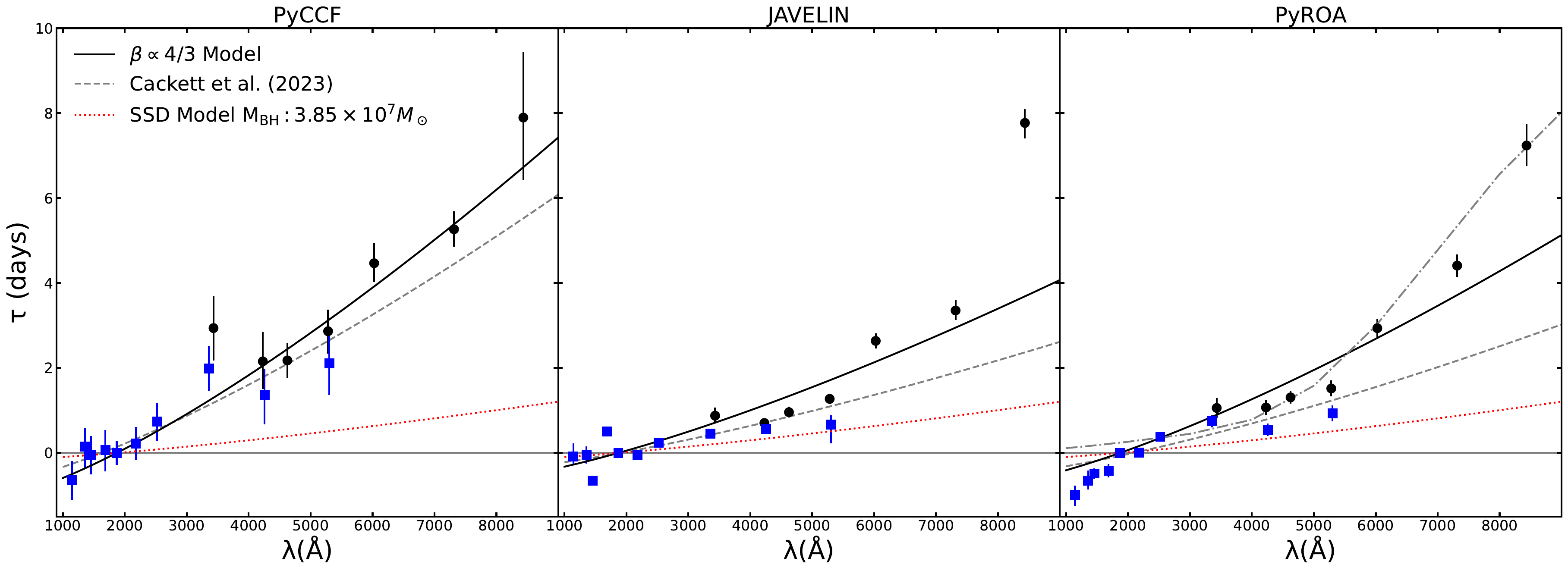}
    \caption{Mrk\,817 lags measured using PyCCF (using $\tau_{\mathrm{cen}}$), JAVELIN, and PyROA. Wavelengths and lags are shown in the rest frame. Blue squares are space-based measurements \citep[HST and Swift;][]{cackett_storm2_IV} and black diamonds are the lags for the ground-based bands. Black curves are the $\beta=4/3$ power-law models (Equation \ref{eqn:powerlaw}) fitted to the data. Gray dashed lines represent fits to the HST and Swift data points as presented by  \citet{cackett_storm2_IV}. Inclusion of the ground-based lags leads to an increase in the normalization factor $\tau_0$, primarily due to the longer lags in the \emph{riz} bands.  Red dotted curves correspond to the expected lag spectrum for lamp-post reprocessing by a Shakura-Sunyaev thin disk  (Equation \ref{eqn:ssdmodel}) for the assumed parameters of Mrk\,817. The dash-dotted line in the rightmost panel is the best-fitting Bowl model as described in Section \ref{sec:bowlmodel}.}
    \label{fig:lagwave}
\end{figure*}

Similar to JAVELIN, PyROA allows the option to fit all light curves simultaneously, but we found that the PyROA runs did not converge properly when applied to the HST, Swift and ground-based filter bands simultaneously. Instead we ran PyROA on light curve pairs using UVW2 and each ground-based band. For consistency, we measured the HST and Swift lags simultaneously to compare to the results in \citet{cackett_storm2_IV}, and our results agreed well (within $1\sigma$). The time delays reported are the median values of the posterior distributions with uncertainties from the 16th to 84th percentile estimated using MCMC methods.

\subsection{Full Campaign Lags}

\begin{deluxetable*}{lcrrcrr}[htbp]
\tablecaption{Full Campaign Lags\label{tab:lags}}
\tablehead{
\colhead{Band/Filter} & \colhead{$\lambda_{\mathrm{cent}}$\,(\AA)} & \colhead{$\tau_{\mathrm{cen}}$\,(days)} & \colhead{$\tau_{\mathrm{peak}}$\,(days)} & \colhead{$r_{\mathrm{max}}$} & \colhead{$\tau_{\mathrm{JAV}}$\,(days)} & \colhead{$\tau_{\mathrm{ROA}}$\,(days)}
}
\startdata
HST-COS 1180 & 1180 & $-0.66^{+0.48}_{-0.46}$ & $-0.68^{+0.24}_{-0.19}$ & $0.88$ & $-0.09^{+0.18}_{-0.32}$ & $-1.02^{+0.27}_{-0.23}$\\
HST-COS 1398 & 1398 & $0.15^{+0.51}_{-0.45}$ & $-0.34^{+0.19}_{-0.19}$ & $0.90$ & $-0.05^{+0.21}_{-0.21}$ & $-0.67^{+0.22}_{-0.24}$\\
HST-COS 1502 & 1502 & $-0.04^{+0.48}_{-0.45}$ & $-0.34^{+0.19}_{-0.19}$ & $0.95$ & $-0.67^{+0.08}_{-0.05}$ & $-0.50^{+0.13}_{-0.13}$\\
HST-COS 1739 & 1739 & $0.07^{+0.52}_{-0.49}$ & $-0.24^{+0.34}_{-0.24}$ & $0.94$ & $-0.52^{+0.07}_{-0.07}$ & $-0.43^{+0.16}_{-0.16}$\\
Swift UVW2 & 1928 & $0.00^{+0.29}_{-0.29}$ & $0.00^{+0.10}_{-0.10}$ & $1.00$ & $0.00^{+0.01}_{-0.01}$ & $0.00^{+0.10}_{-0.10}$ \\
Swift UVM2 & 2246 & $0.23^{+0.40}_{-0.40}$ & $-0.19^{+0.53}_{-0.15}$ & $0.99$ & $-0.05^{+0.08}_{-0.08}$ & $0.01^{+0.08}_{-0.08}$\\
Swift UVW1 & 2600 & $0.76^{+0.46}_{-0.46}$ & $0.39^{+0.10}_{-0.15}$ & $0.98$ & $0.25^{+0.07}_{-0.07}$ & $0.39^{+0.10}_{-0.10}$\\
Swift U & 3465 & $2.05^{+0.55}_{-0.55}$ & $0.48^{+0.10}_{-0.15}$ & $0.97$ & $0.47^{+0.07}_{-0.07}$ & $0.77^{+0.15}_{-0.15}$\\
Swift B & 4392 & $1.41^{+0.72}_{-0.62}$ & $0.48^{+0.29}_{-0.19}$ & $0.95$ & $0.58^{+0.08}_{-0.08}$ & $0.56^{+0.15}_{-0.15}$\\
Swift V & 5468 & $2.18^{+0.77}_{-0.72}$ & $0.68^{+1.36}_{-0.34}$ & $0.94$ & $0.69^{+0.46}_{-0.22}$ & $0.96^{+0.19}_{-0.19}$\\
\hline
$u$ & 3540 & $2.94^{+0.77}_{-0.76}$ & $0.90^{+1.73}_{-2.33}$ & $0.95$ & $0.88^{+0.19}_{-0.19}$ & $1.06^{+0.24}_{-0.23}$\\
$B$ & 4361 & $2.16^{+0.65}_{-0.69}$ & $1.00^{+1.02}_{-1.62}$ & $0.96$ & $0.71^{+0.11}_{-0.11}$ & $1.07^{+0.18}_{-0.18}$\\
$g$ & 4770 & $2.18^{+0.41}_{-0.42}$ & $1.10^{+0.55}_{-1.45}$ & $0.96$ & $0.96^{+0.13}_{-0.13}$ & $1.31^{+0.15}_{-0.15}$\\
$V$ & 5448 & $2.87^{+0.53}_{-0.51}$ & $1.70^{+0.86}_{-1.86}$ & $0.96$ & $1.27^{+0.12}_{-0.10}$ & $1.52^{+0.18}_{-0.19}$\\
$r$ & 6215 & $4.47^{+0.44}_{-0.48}$ & $2.30^{+1.61}_{-2.71}$ & $0.95$ & $2.64^{+0.18}_{-0.18}$ & $2.94^{+0.21}_{-0.21}$\\
$i$ & 7545 & $5.27^{+0.41}_{-0.43}$ & $3.80^{+0.93}_{-2.23}$ & $0.94$ & $3.35^{+0.22}_{-0.24}$ & $4.41^{+0.26}_{-0.27}$\\
$z$ & 8700 & $7.90^{+1.47}_{-1.55}$ & $5.10^{+1.25}_{-4.25}$ & $0.88$ & $7.71^{+0.36}_{-0.32}$ & $7.24^{+0.48}_{-0.51}$\\
\enddata
\tablecomments{Reverberation lags for the full STORM\,2 campaign for Mrk\,817, measured by the ICCF, JAVELIN, and PyROA methods. Central wavelengths are in the observed frame and lags are in the rest frame of Mrk 817. HST and Swift lags reported here are reproduced from Table 2 of \citet{cackett_storm2_IV}.}
\end{deluxetable*}

The lags for the full campaign are reported in Table\,\ref{tab:lags} for all three measurement methods and shown as  lag-wavelength plots in Figure\,\ref{fig:lagwave}. For completeness, Table \ref{tab:lags} also includes the lags for the HST and Swift bands presented in Paper IV. All methods show lags increasing with wavelength in agreement with previous STORM\,2 results \citep{Kara_2021,cackett_storm2_IV} and consistent with most broad-band continuum RM studies for other AGN. Consistent with earlier results from \citet{cackett_storm2_IV}, the PyCCF measurements exhibit a distinct excess lag in the U/\emph{u} bands, but the JAVELIN and PyROA lags do not show clear evidence for a $u$-band excess.

Comparing the lags obtained from the three methods, we find that PyCCF generally yields longer lags ($\tau_\mathrm{cen}$) than JAVELIN or PyROA, as previously found by \citet{cackett_storm2_IV} for the HST and Swift bands. This is a consequence of using the CCF centroid (rather than the CCF peak) to quantify the lag, since the CCFs have asymmetric shapes with excess power at lags longer than the peak (Figure \ref{fig:javmodels}). The asymmetry in the CCFs can result from the shape of the disk reprocessing transfer functions created by the radial temperature profile of the disk \citep{Cackett2007}, and DC and line emission from the BLR will contribute a tail extending to lags well beyond the peak. The CCF peak lags ($\tau_\mathrm{peak}$) for the optical bands are all smaller than the corresponding $\tau_\mathrm{cen}$ values, and are closer to the lags obtained with JAVELIN and PyROA (except for the $z$ band). We also find that PyCCF yields larger lag uncertainties than the other methods, as is typically the case in prior reverberation studies \citep{Fausnaugh_2016,Edelson2019,Yu_uncertainties,mrk142,2024A&AT...34..221G}.

The wavelength range covered by the Swift and ground-based filters overlaps across the UBV bands (see Figure\,\ref{fig:specfilter}). We find that the ground-based lags are consistently slightly longer than the corresponding Swift-based lags for each overlapping filter and for all three lag measurement methods. The mean lag differences between the overlapping space and ground-based bands are $\Delta\tau = 0.78$ days for lags measured with PyCCF, $\Delta\tau = 0.37$ days with JAVELIN, and $\Delta\tau = 0.45$ days with PyROA. These differences are within the $1\sigma$ uncertainties for PyCCF, and within $3\sigma$ for JAVELIN and PyROA (where both the lag differences and their uncertainties are smaller than for PyCCF). The reason for these differences is not clear, and may be due to the differences between the ground-based and Swift filter passbands and/or the different cadence of the ground-based light curves. We carried out tests to check whether this discrepancy might be caused by the different time spans of the ground-based and space-based light curves by truncating the ground-based light curves to match the end date of the Swift light curves at THJD 59640 and re-measuring the CCF lags. We found that the revised ground-based lags remained consistent but were slightly longer compared to the lags of the full-duration light curves, indicating that the discrepancy with the Swift UBV lags is not the result of the longer duration of the ground-based light curves.

\added{To test whether the lags are significantly impacted by the final portion of the campaign after THJD 9600, during which there are two large gaps in the Swift light curves, we remeasured the lags using versions of the light curves that were truncated at THJD 9600. We found that the measured lags changed by less than $1.5\sigma$ compared with the full-campaign lags listed in Table \ref{tab:lags}, indicating that the results are largely insensitive to the inclusion of the data from that interval. The only exception was for the \emph{z} band, where the Javelin results for the truncated light were somewhat anomalous, showing substantial power at (unphysical) negative lag values and yielding a median lag of $\sim$4.8 days, about $8\sigma$ lower than the lag obtained for the full-duration light curves.}

To evaluate the trend of lag with wavelength we fit an offset power-law model to the lag spectrum as done in  many other works \citep[e.g.,][]{Fausnaugh_2016,Edelson2019}. The model has the form
\begin{equation}
    \tau = \tau_0[(\lambda/\lambda_0)^\beta - 1]
    \label{eqn:powerlaw}
\end{equation}
where $\tau$ is the rest-frame lag, $\lambda_0 = 1928/(1+z)$\,\AA\ is the reference wavelength (corresponding to the rest-frame wavelength at the center of the UVW2 band), $\beta$ is fixed to $4/3$ corresponding to a basic disk reprocessing model, and the free parameter $\tau_0$ sets the normalization of the lag spectrum. 

The fits to the lag spectra measured with each method yield  $\tau_0 = 1.04 \pm 0.05$ days with $\chi^2_\textrm{dof}=1.03$ for PyCCF (using $\tau_\mathrm{cen}$), $\tau_0 = 0.59 \pm 0.09$ days with $\chi^2/\textrm{dof}=0.49$ for PyCCF (using $\tau_\mathrm{peak}$), $\tau_0 = 0.57 \pm 0.08$ days with $\chi^2_\textrm{dof}=23.9$ for JAVELIN, and $\tau_0 = 0.72 \pm 0.11$ days with $\chi^2_\textrm{dof}=28.3$ for PyROA. All $\chi^2_\textrm{dof}$ are reported for $15$ degrees of freedom. While there are 17 continuum bands (including HST, Swift, and ground-based bands; see Table\,\ref{tab:lags}) and one free parameter in the fit, the model  of Equation \ref{eqn:powerlaw} is constructed to pass precisely through the UVW2 data point having zero lag, so the UVW2 point does not count toward the number of degrees of freedom in the fit. 

For comparison, we have also included in Figure\,\ref{fig:lagwave} the power-law model from \citet{cackett_storm2_IV} that was fit to just the HST and Swift data points. \citet{cackett_storm2_IV} used a model of the form $\tau=\tau_0[(\lambda/\lambda_0)^\beta -y_0]$, with $y_0$ as a free parameter rather than fixed as in our Equation \ref{eqn:powerlaw}, so the fit is not forced to coincide with the UVW2 data point (but the fitted curves fall very close to the UVW2 point nevertheless). For each lag measurement method, we find that including the ground-based bands yields a larger normalization $\tau_0$, due to the longer lags measured for the ground-based \emph{uBV} bands compared with their Swift equivalents, and the significantly longer lags of the \emph{riz} bands. Our $\tau_0$ values are larger by factors of 1.25, 1.49, and 1.78 than the values obtained by \citet{cackett_storm2_IV} from fits to the space-based points, for $\tau_\mathrm{cen}$, JAVELIN, and PyROA, respectively. 

To compare our results with predictions for lamp-post reprocessing by a thin disk, we adopt the model described by \citet{Fausnaugh_2016}. This model assumes a geometrically thin, optically thick accretion disk that is irradiated by an X-ray corona. The standard thin disk model has a temperature profile of $T(R)\propto R^{-3/4}$. Assuming a relationship between characteristic emitted wavelength $\lambda$ and disk temperature of the form $\lambda=Xhc/kT$, while $R=c\tau$ (the time lags converted to light-travel distance) then yields the $\tau\propto\lambda^{4/3}$ lag-wavelength relation. Here, $X$ is a factor that specifies the relation between $T(R)$ and $\lambda$ for a given $R$, accounting for the fact that multiple radii contribute to the emission at any given wavelength. Specifically, we use $X=2.49$ corresponding to a flux-weighted mean radius for a disk that emits locally as a blackbody \citep[see Equation 10 of][]{Fausnaugh_2016}. The predicted value of $\tau_0$ \citep[Equation 12 from][]{Fausnaugh_2016} is 
\begin{equation}
    \tau_0 = \frac{1}{c} \left(X\frac{k\lambda_0}{hc}\right)^{4/3}\left[\left(\frac{GM}{8\pi\sigma}\right)  \left(\frac{L_\textrm{Edd}}{\eta c^2}\right)(3+\kappa)\,\dot{m_\textrm{E}}\right]^{1/3}.
    \label{eqn:ssdmodel}
\end{equation}

We adopt $\eta = 0.1$ for the radiative efficiency, and $\kappa = 1$ for the ratio of external to internal (viscous) heating of the disk. For the Eddington ratio we use $\dot{m}=0.1$ following Paper X, since this value is based on the observed disk luminosity (at 5100 \AA) corrected for the estimated contribution of DC emission. These assumptions yield $\tau_0=0.169$ days which is much smaller than the values obtained from fits to the observed lag spectrum, as illustrated in Figure \ref{fig:lagwave}.  For PyCCF the lags are a factor of $6.2$ times longer, for JAVELIN a factor of $3.0$ times longer, and for PyROA a factor of $4.3$ times longer than the disk reprocessing model prediction. These discrepancy factors are similar to values found for other AGN \citep[e.g.,][]{Fausnaugh_2016,Edelson2019} as well as for lensed quasars; for example, \citet{Morgan2010} found quasar disk sizes at 2500 \AA\ to be $\sim$4 times larger than predicted by standard thin-disk models. However, it should be emphasized that the magnitude of the discrepancy depends on multiple uncertain factors including the assumed values of $X$, $\eta$, and Eddington ratio in Equation \ref{eqn:ssdmodel}. If $X=3.37$ were adopted, representing the response weighted by variable irradiation as noted by \citet{Fausnaugh_2016}, the size discrepancy factor would be reduced correspondingly.  

While PyCCF generally produces longer lags $(\tau_\mathrm{cen}$) than the other methods, the $z$-band lags are consistent (within their $1\sigma$ uncertainties) for all three measurement methods, at 7.2-7.9 days. The PyCCF lag spectrum is reasonably well fit by the power-law model across all bands from the far-UV through $z$, but for JAVELIN and PyROA the $z$-band points are notable outliers above the model fits, and contribute significantly to the high $\chi^2$ values. \added{Previously, \citet{hagai_storm2_X} proposed that the long \emph{z}-band lag might be due to a contribution from dust in the inner torus at $T \approx 1600$ K. However, models of the near-IR spectral energy distribution by \cite{landt_storm2_IR} (observed when Mrk 817 was close to its median flux during the campaign) found a somewhat cooler dust temperature of 1430 K, implying that dust contributes about 8\% of the nuclear \emph{z}-band flux. This would be too small to have a strong effect on the overall lag measurements, although its presence could still potentially be detected through other analysis methods \citep[e.g.,][]{Honig_2014}.} The possible dust contribution at long wavelengths can be further explored in future work by applying methods such as MEMEcho \citep{horne_storm1_IX} to examine the response function shape and search for evidence of multiple components in the lag distribution.

\subsection{Lags of Short-Duration Epochs} \label{sec:epochlag}

\startlongtable
\begin{deluxetable*}{lrrcrr}
\tablecaption{Lag Measurements for Individual Epochs \label{large_lag_table}}
\tablehead{\\[-1.2cm]}
\startdata
&\multicolumn{5}{c}{Epoch 1}\\
\hline
Filter/Band&$\tau_\textrm{cen}$&$\tau_\textrm{peak}$&$r_\textrm{max}$&$\tau_\textrm{JAV}$&$\tau_\textrm{ROA}$\\
\hline
HST 1180 &$-1.17^{+0.47}_{-0.43}$ & $-0.78^{+0.68}_{-0.04}$ & $0.80$&$-1.33^{+0.24}_{-0.17}$&$-0.77^{+0.14}_{-0.14}$\\ 
HST 1398 &$-0.67^{+0.46}_{-0.45}$ & $-0.68^{+0.28}_{-0.20}$ & $0.85$&$-1.56^{+0.10}_{-0.13}$&$-0.60^{+0.16}_{-0.16}$\\
HST 1502 &$-0.30^{+0.41}_{-0.39}$ & $-0.39^{+0.11}_{-0.38}$ &$0.93$&$-0.20^{+0.13}_{-0.11}$&$-0.38^{+0.16}_{-0.15}$\\
HST 1739 &$0.59^{+0.53}_{-0.48}$  & $-0.58^{+0.30}_{-0.19}$ &$0.91$&$-0.72^{+0.20}_{-0.20}$&$-0.71^{+0.25}_{-0.25}$\\
Swift UVM2&$0.23^{+0.58}_{-0.64}$  & $0.23^{+0.16}_{-0.61}$ &$0.97$&$0.00^{+0.14}_{-0.15}$&$0.16^{+0.16}_{-0.16}$\\
Swift UVW1&$0.47^{+0.60}_{-0.72}$  & $0.47^{+0.21}_{-0.76}$ & $0.95$&$0.50^{+0.27}_{-0.27}$&$0.49^{+0.28}_{-0.28}$\\
Swift U&$1.31^{+0.69}_{-0.72}$  & $1.31^{+0.14}_{-1.51}$ & $0.88$&$1.92^{+0.45}_{-0.45}$&$1.53^{+0.46}_{-0.45}$\\
Swift B&$1.46^{+1.00}_{-1.12}$  & $1.46^{+0.10}_{-1.56}$ & $0.83$&$1.80^{+0.79}_{-0.87}$&$2.11^{+0.59}_{-0.58}$\\
Swift V &$2.84^{+1.12}_{-0.96}$  & $2.84^{+0.22}_{-2.64}$ & $0.80$&$3.54^{+0.67}_{-0.63}$&$3.81^{+0.60}_{-0.56}$\\
Ground \emph{u} &$1.80^{+0.79}_{-0.60}$  & $0.97^{+0.43}_{-1.60}$ & $0.85$&$1.85^{+0.08}_{-0.11}$&$1.81^{+0.43}_{-0.45}$\\
Ground \emph{B} &$1.66^{+0.74}_{-0.63}$  & $1.26^{+0.38}_{-1.07}$ & $0.86$&$1.65^{+0.71}_{-0.71}$&$1.62^{+0.39}_{-0.37}$\\
Ground \emph{g} &$2.71^{+0.48}_{-0.55}$  & $2.04^{+0.19}_{-1.26}$ & $0.87$&$1.74^{+0.25}_{-0.25}$&$2.04^{+0.35}_{-0.36}$\\
Ground \emph{V} &$2.97^{+0.68}_{-0.64}$  & $1.84^{+0.25}_{-1.99}$ & $0.81$&$3.46^{+0.70}_{-0.63}$&$2.56^{+0.39}_{-0.40}$\\
Ground \emph{r} &$4.50^{+0.53}_{-0.69}$  & $4.17^{+1.03}_{-1.68}$ & $0.78$&$3.85^{+0.29}_{-0.31}$&$3.64^{+0.45}_{-0.45}$\\
Ground \emph{i} &$5.52^{+0.60}_{-0.68}$  & $4.65^{+1.18}_{-1.63}$ & $0.80$&$5.20^{+0.36}_{-0.33}$&$4.98^{+0.43}_{-0.41}$\\
Ground \emph{z} &$7.60^{+2.78}_{-1.81}$  & $5.82^{+4.02}_{-3.74}$ & $0.64$&$8.96^{+1.01}_{-1.02}$&$9.83^{+0.94}_{-0.97}$\\
\hline
&\multicolumn{5}{c}{Epoch 2}\\
\hline
&$\tau_\textrm{cen}$&$\tau_\textrm{peak}$&$r_\textrm{max}$&$\tau_\textrm{JAV}$&$\tau_\textrm{ROA}$\\
\hline
HST 1180 &$-0.68^{+0.56}_{-0.47}$ & $-0.68^{+0.49}_{-0.19}$ & $0.68$&$-0.46^{+0.96}_{-0.10}$&$-0.65^{+0.14}_{-0.14}$\\
HST 1398 &$-0.31^{+0.53}_{-0.52}$ & $-0.48^{+0.71}_{-0.45}$ & $0.70$&$-0.44^{+0.08}_{-0.98}$&$-0.32^{+0.16}_{-0.16}$\\
HST 1502  &$-0.38^{+0.60}_{-0.77}$ & $-0.48^{+0.96}_{-0.39}$ & $0.79$&$-0.49^{+0.08}_{-0.09}$&$-0.46^{+0.16}_{-0.15}$\\
HST 1739 &$-0.23^{+0.66}_{-0.89}$ & $-0.29^{+1.29}_{-0.55}$ &$0.82$&$-0.20^{+0.47}_{-0.29}$&$-0.11^{+0.25}_{-0.25}$\\
Swift UVM2 &$0.19^{+0.33}_{-0.34}$  & $0.19^{+0.20}_{-0.48}$ & $0.96$&$-0.05^{+0.20}_{-0.13}$&$-0.05^{+0.12}_{-0.11}$\\
Swift UVW1 &$0.78^{+0.36}_{-0.37}$  & $0.29^{+0.19}_{-0.48}$ & $0.93$&$0.27^{+0.14}_{-0.14}$&$0.48^{+0.18}_{-0.17}$\\
Swift U &$0.77^{+0.48}_{-0.46}$  & $0.39^{+0.18}_{-0.67}$ & $0.91$&$0.23^{+0.20}_{-0.20}$&$0.47^{+0.21}_{-0.21}$\\
Swift B &$0.87^{+0.47}_{-0.44}$  & $0.68^{+0.59}_{-0.48}$ & $0.85$&$0.72^{+0.30}_{-0.30}$&$0.84^{+0.21}_{-0.21}$\\
Swift V &$0.99^{+0.65}_{-0.66}$  & $0.78^{+0.95}_{-1.18}$ & $0.83$&$0.84^{+0.47}_{-0.44}$&$0.80^{+0.27}_{-0.29}$\\
Ground \emph{u} &$0.09^{+0.71}_{-0.62}$  & $0.19^{+0.77}_{-0.68}$ & $0.91$&$0.35^{+0.24}_{-0.49}$&$0.29^{+0.27}_{-0.28}$\\
Ground \emph{B} &$0.62^{+0.71}_{-0.70}$  & $0.97^{+0.74}_{-0.23}$ &  $0.93$&$0.79^{+0.26}_{-0.25}$&$1.01^{+0.23}_{-0.23}$\\
Ground \emph{g} &$1.20^{+0.28}_{-0.27}$  & $1.07^{+0.26}_{-0.42}$ & $0.96$&$1.04^{+0.17}_{-0.16}$&$1.16^{+0.15}_{-0.16}$\\
Ground \emph{V} &$1.44^{+0.50}_{-0.53}$  & $1.36^{+0.40}_{-0.57}$ & $0.94$&$0.97^{+0.38}_{-0.39}$&$1.58^{+0.20}_{-0.20}$\\
Ground \emph{r} &$2.47^{+0.42}_{-0.44}$  & $1.94^{+0.24}_{-1.21}$ & $0.93$&$2.11^{+0.21}_{-0.23}$&$2.22^{+0.25}_{-0.25}$\\
Ground \emph{i} &$3.41^{+0.41}_{-0.43}$  & $2.91^{+0.16}_{-1.30}$ & $0.92$&$3.09^{+0.19}_{-0.20}$&$3.21^{+0.24}_{-0.24}$\\
Ground \emph{z} &$4.41^{+0.98}_{-1.20}$  & $3.97^{+0.53}_{-1.22}$ & $0.89$&$4.72^{+0.52}_{-0.61}$&$4.28^{+0.32}_{-0.33}$\\
\hline
&\multicolumn{5}{c}{Epoch 3}\\
\hline
&$\tau_\textrm{cen}$&$\tau_\textrm{peak}$&$r_\textrm{max}$&$\tau_\textrm{JAV}$&$\tau_\textrm{ROA}$\\
\hline
HST 1180 &$0.31^{+0.70}_{-0.95}$  & $-0.48^{+0.02}_{-1.28}$ & $0.96$&$-0.66^{+0.13}_{-0.05}$&$-0.94^{+0.13}_{-0.12}$ \\
HST 1398 &$1.02^{+0.63}_{-0.91}$  & $0.10^{+0.44}_{-1.51}$ & $0.96$&$-0.73^{+0.10}_{-0.07}$&$-0.56^{+0.11}_{-0.11}$\\
HST 1502 &$0.63^{+0.58}_{-0.73}$  & $-0.19^{+0.15}_{-1.22}$ &  $0.96$&$-0.78^{+0.07}_{-0.08}$&$-0.70^{+0.12}_{-0.11}$\\
HST 1739 &$0.64^{+0.75}_{-0.87}$  & $0.48^{+0.38}_{-1.54}$ &$0.96$&$ -0.06^{+0.23}_{-0.23}$&$-0.04^{+0.18}_{-0.17}$\\
Swift UVM2&$0.41^{+0.60}_{-0.59}$  & $-0.29^{+0.02}_{-0.80}$ & $0.98$&$-0.11^{+0.13}_{-0.15}$&$-0.07^{+0.13}_{-0.13}$\\
Swift UVW1&$0.48^{+0.52}_{-0.50}$  & $0.29^{+0.00}_{-0.67}$ & $0.98$&$0.12^{+0.10}_{-0.11}$&$0.29^{+0.11}_{-0.12}$\\
Swift U   &$1.84^{+0.53}_{-0.53}$  & $0.48^{+1.07}_{-1.65}$ & $0.97$&$0.56^{+0.22}_{-0.34}$&$0.66^{+0.18}_{-0.17}$\\
Swift B   &$0.59^{+0.57}_{-0.53}$  & $0.39^{+0.09}_{-0.88}$ & $0.96$&$0.27^{+0.18}_{-0.19}$&$0.34^{+0.18}_{-0.16}$\\
Swift V   &$1.58^{+0.75}_{-0.72}$  & $0.58^{+0.65}_{-1.97}$ & $0.92$&$0.97^{+0.49}_{-0.41}$&$0.61^{+0.25}_{-0.26}$\\
Ground \emph{u} &$3.59^{+1.03}_{-1.09}$  & $1.94^{+0.78}_{-2.82}$ & $0.92$&$ 1.29^{+0.36}_{-0.39}$&$1.26^{+0.29}_{-0.28}$\\
Ground \emph{B} &$1.10^{+0.98}_{-1.07}$  & $0.29^{+0.33}_{-1.30}$ & $0.96$&$0.36^{+0.34}_{-0.20}$&$0.87^{+0.56}_{-0.50}$\\
Ground \emph{g} &$1.64^{+0.49}_{-0.44}$  & $0.97^{+0.11}_{-0.96}$ & $0.97$&$0.97^{+0.15}_{-0.15}$&$1.11^{+0.15}_{-0.14}$\\
Ground \emph{V} &$3.36^{+0.61}_{-0.72}$  & $2.13^{+0.54}_{-2.00}$ & $0.92$&$1.85^{+0.10}_{-0.06}$&$1.08^{+0.56}_{-0.60}$\\
Ground \emph{r} &$4.26^{+0.47}_{-0.48}$  & $2.81^{+0.28}_{-2.13}$ & $0.95$&$4.12^{+0.30}_{-0.30}$&$3.63^{+0.25}_{-0.25}$\\
Ground \emph{i} &$4.98^{+0.47}_{-0.52}$  & $4.75^{+1.42}_{-1.68}$ & $0.92$&$ 4.60^{+0.27}_{-0.27}$&$4.89^{+0.36}_{-0.37}$\\
Ground \emph{z} &$6.48^{+2.55}_{-2.92}$  & $6.50^{+3.02}_{-2.41}$ & $0.82$&$5.92^{+0.64}_{-0.63}$&$6.00^{+0.69}_{-0.72}$\\
\enddata
\tablecomments{All lags reported are in the AGN rest frame in units of days.}
\end{deluxetable*}


The long duration of the STORM 2 campaign and the high S/N of the light curves provides an opportunity to measure the broad-band lags over shorter subsets of the full monitoring campaign and to test for changes in the lags that might be correlated with variations in AGN luminosity or obscuration.

The frequency-resolved lag analysis in Paper VII \citep{lewin_storm2_VII} split the light curves into three segments (denoted as ``epochs'') of equal duration to investigate whether there are changes in lags. The first and third epochs had higher time-averaged X-ray obscuration ($\bar{N}_\mathrm{H} = 12.6\times10^{22}$ cm$^{-2}$ and $12.9\times10^{22}$ cm$^{-2}$) compared to the second epoch ($\bar{N}_\mathrm{H} = 6.1\times10^{22}$ cm$^{-2}$), based on X-ray measurements from Paper III \citep{Partington_2023_STORM2_III}. \citet{lewin_storm2_VII} found that the lags corresponding to low-frequency flux variations (at frequencies of $0.014 - 0.048$ day$^{-1}$) were longer during the first and third epochs, when the obscuration was higher, and the amplitude of the lag spectrum (i.e., the inferred value of $\tau_0$ for lags in this frequency range) dropped by nearly a factor of two during Epoch 2 compared with Epochs 1 and 3.

Here, we follow up on this investigation and measure time delays for the same three epochs, to test whether the standard methods for lag determination yield similar results for the variation in lags across the duration of the campaign.  We adopt THJD 9317 and 9457 as the boundaries between Epochs 1 and 2, and between Epochs 2 and 3, following \citet{lewin_storm2_VII}.
These epoch boundaries (see Figure \ref{fig:javmodels}) were chosen in order to split the Swift light curves into segments of about 140 days each. The third epoch ends at THJD 9600, excluding the final portion of the light curves,  since only six Swift observations were obtained after that date, leaving a large gap in the Swift coverage after THJD 9600. We select these same epoch boundaries for direct comparison with the results of \citet{lewin_storm2_VII}. 

For each epoch, we measured the lags of each band relative to UVW2 following the same procedures described in Section \ref{sec:lags}, using PyCCF, JAVELIN, and PyROA. The derived lags for each epoch, band, and method are listed in Table \ref{large_lag_table}. A notable change across the three epochs is seen in the $r_\mathrm{max}$ values, where the optical bands show consistently lower correlations with the UVW2 light curve during Epoch\,1 than during Epochs\,2 and 3. These differences stem from the wavelength-dependent divergence in the light curve shapes that temporarily occurred near the start of the campaign (during THJD $\approx9220-9320$), as highlighted by \citet{cackett_storm2_IV} from the HST and Swift data. 

\begin{figure*}
    \centering
    \includegraphics[width=\linewidth]{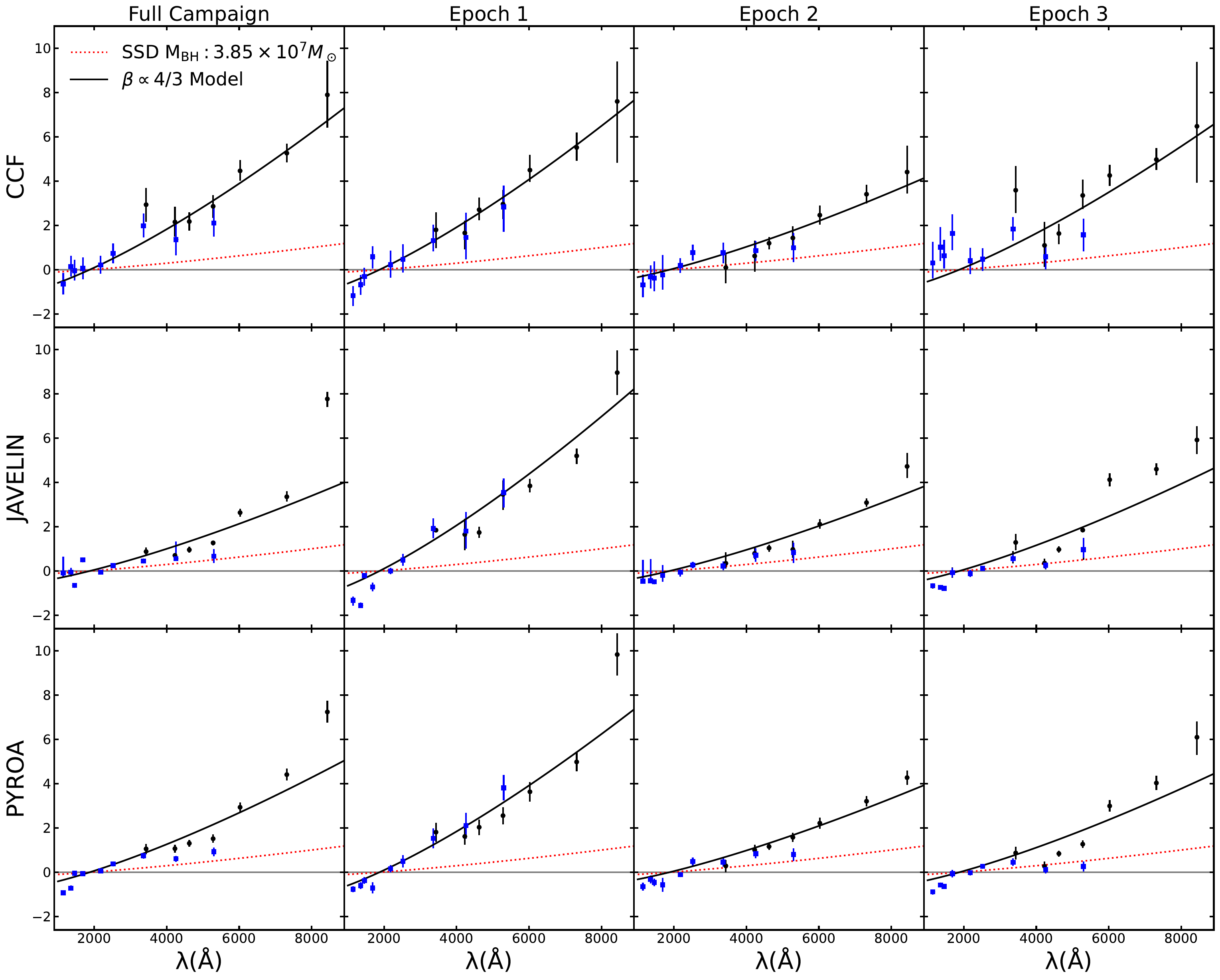}
    \caption{Lag-wavelength relations for the full campaign and for the three epochs, for each lag measurement method. Model curves are as described in Figure\,\ref{fig:lagwave}. Blue points are the space-based (HST and Swift) lag measurements, while black points are all ground-based.}
    \label{fig:epoch_lag_wave}
\end{figure*}

The lag spectrum for each epoch was then fit with the power-law model of Equation \ref{eqn:powerlaw}. The epoch lags are illustrated in Figure \ref{fig:epoch_lag_wave} and the $\tau_0$ and $\chi^2_\textrm{dof}$ values from each fit are listed in Table \ref{epochlagstab}. All of our lag measurement methods show a factor of $\sim$2 drop in $\tau_0$ from Epoch 1 to Epoch 2, as previously found by \citet{lewin_storm2_VII}, although our derived values of $\tau_0$ are systematically longer than those measured with the frequency-resolved technique. 
For Epoch 3, \citet{lewin_storm2_VII} found that $\tau_0$ rose again to nearly the same value as in Epoch 1. We find similar behavior for our $\tau_\mathrm{cen}$ measurement, while in contrast, our $\tau_\mathrm{peak}$, JAVELIN, and PyROA-derived values of $\tau_0$ remain low in Epoch 3, consistent with their values during Epoch 2. Thus, the overall trend across the three epochs for the low-frequency lags from \citet{lewin_storm2_VII} appears most similar to what we find using $\tau_\mathrm{cen}$.

\begin{deluxetable*}{lcccccccc}[]
\tablecaption{Lag Spectrum Fitting Results \label{epochlagstab}}
\tablehead{
\colhead{Method} & \colhead{$\tau_0$(Full Campaign)} & \colhead{$\chi^2_\textrm{dof}$} &\colhead{$\tau_0$(Epoch 1)} & \colhead{$\chi^2_\textrm{dof}$} & \colhead{$\tau_0$(Epoch 2)}  &  \colhead{$\chi^2_\textrm{dof}$}& \colhead{$\tau_0$(Epoch 3)} & \colhead{$\chi^2_\textrm{dof}$}
}
\startdata
PyCCF ($\tau_{\textrm{cen}}$) & $1.04\pm0.05$ & $1.03$ & $1.09\pm0.05$ & $0.46$ & $0.59\pm0.04$ &  $0.54$  & $0.94\pm0.09$ & $2.08$\\
PyCCF ($\tau_{\textrm{peak}}$) & $0.59\pm 0.09$ & $0.49$ & $0.92\pm0.09$ & $0.59$ & $0.51\pm0.03$ & $0.32$ & $0.51\pm0.09$ & $0.62$\\
JAVELIN & $0.57\pm 0.08 $& $23.9$ & $1.17\pm 0.12$ & $11.41$ & $0.54\pm0.04$ &$2.65$& $0.66\pm0.08$ & $16.12$\\
PyROA & $0.72\pm0.11$ & $28.3$ & $1.05\pm0.07$&$2.10$ & $0.56\pm0.04$&$2.51$ & $0.63\pm0.09$ &$12.18$\\
\enddata
\tablecomments{All $\tau_0$ values are in units of days, and are obtained from fitting Equation \ref{eqn:powerlaw} to the rest-frame lags as shown in Figure \ref{fig:epoch_lag_wave}. The reduced $\chi^2$ values are for $15$ degrees of freedom.} 
\end{deluxetable*}

All epochs exhibit a disk size discrepancy in comparison with the reprocessing model of Equation \ref{eqn:ssdmodel}, with the largest discrepancies found during the first epoch. For the PyCCF measurements, the magnitude of the discrepancies for each epoch (i.e., the ratio of observed to predicted values of $\tau_0$) are a factor of $6.5$ for Epoch 1, $3.5$ for Epoch 2, and $5.5$ for Epoch 3. For JAVELIN the corresponding factors are $6.9$, $3.2$, and $3.9$ for Epochs 1, 2, and 3, respectively, while for PyROA we obtain factors of $6.2$, $3.3$, and $3.8$ for the three epochs. 

\added{We have not measured lags for the brief light curve segment after the end of Epoch 3 (THJD 9600-9700) since only a few Swift observations were taken during this period prior to the end of the Swift STORM 2 observing program on THJD 9635. However, Swift monitoring did continue after THJD 9635 as part of the extended campaign, and the Swift and ground-based lags for this time period will be examined in future work.}

\section{Flux-Flux Analysis}
\label{sec:flxflx}
\subsection{Full Campaign}
The flux-flux analysis method \citep{Winkler1992,fairall9juan} decomposes the observed light curves into variable and constant components, with the goal of recovering the broad-band spectral shape of the time-variable AGN emission. By construction, the flux-flux method assumes that the variable component has a constant spectral shape as it varies in luminosity. \citet{cackett_storm2_IV} carried out a flux-flux analysis of the STORM 2 light curves from HST and Swift, finding that the rms variability spectrum follows a spectral shape of $f_\nu \propto \lambda^{-1/3}$ from the UV through V bands. Here, we extend this flux-flux analysis by adding the ground-based light curves and by examining the flux-flux spectral components in the three epochs described in Section \ref{sec:epochlag}. To match the duration of the space-based light curves, we removed the final 35 days from the ground-based light curves for this analysis. The light curves are first corrected for a Galactic reddening of $E(B-V)=0.022$ mag using a \citet{Cardellireddening} extinction law with $R_V = 3.1$.  

Following the same methods described by \citet{cackett_storm2_IV}, we fit the data with a dimensionless model light curve $X(t)$ that is shifted by an additive offset $A_\nu (\lambda)$ and scaled by $S_\nu(\lambda)$ to fit each band's light curve, where $S_\nu(\lambda)$ and $A_\nu (\lambda)$ are free parameters, and $X(t)$ is constrained to have $\langle X \rangle = 0$ and $\langle X^2 \rangle = 1$. Then, the resulting model light curve for wavelength $\lambda$  is 
\begin{equation}
f_\nu(\lambda, t) = A_\nu(\lambda) + S_\nu(\lambda)X(t). 
\label{eqn:fluxflux}
\end{equation}

\begin{figure}
    \centering
    \includegraphics[width=\linewidth]{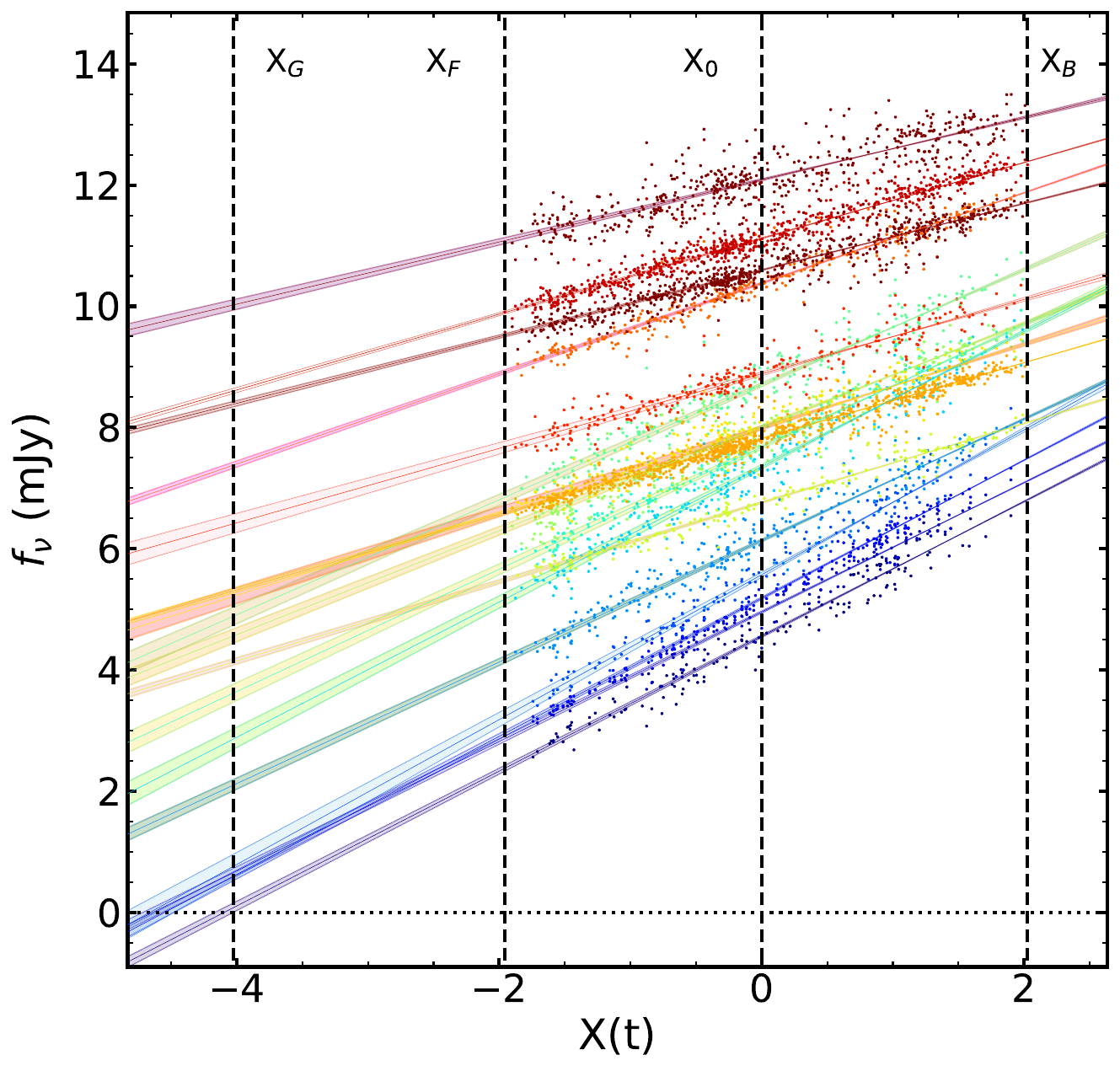}
    \caption{Flux-flux relations for the full campaign duration for each UV and optical band, with the 1180 \AA\ band at the bottom (dark violet) and $z$ at the top (dark red). The data for each filter-band show the variability range of the light curves, and the lines show the fit of Equation \ref{eqn:fluxflux} to each band along with its $1\sigma$ uncertainty range. The quantity $X_G$ is the value of $X$ that is used to establish the constant component flux for each band, as described in the text, and the slope of the fitted line for each filter-band gives the variability amplitude parameter $S_\nu(\lambda)$ for that band. $X_F$ and $X_B$ denote the faintest and brightest states of the light curves, and $X_0$ corresponds to the mean flux level. }
    \label{fig:fluxflux}
\end{figure}

The resulting flux-flux light curve relations are shown in Figure \ref{fig:fluxflux}. The best-fit scale factors $S_\nu(\lambda)$ give the rms variability amplitude for each wavelength band. To obtain the constant component of the spectrum, we find the value $X_G$ where the lower error envelope on the flux-flux relation for the shortest-wavelength band (1180 \AA) reaches zero flux. Then, for each band the value of $f_\nu$ at $X=X_G$ is taken to be the constant component flux for that band.

\begin{deluxetable*}{lCCCCC}[t]
\tablecaption{Mrk\,817 Broad-Band Flux-Flux Spectral Data}
\tablehead{\colhead{Filter Band} & \colhead{Mean $f_\nu$} & \colhead{Bright $f_\nu$} & \colhead{Faint $f_\nu$} & \colhead{$S_\nu$} &\colhead{$A_\nu$} }
\startdata
1180 \AA & $4.61$ & $6.50\pm0.07$ & $2.47\pm0.06$ & $1.11\pm0.01$ & $0.08\pm0.04$\\
1398 \AA & $5.00$ & $6.48\pm0.06$ & $2.47\pm0.05$ & $1.07\pm0.01$ & $0.66\pm0.03$\\
1502 \AA& $5.23$ & $6.73\pm0.06$ & $2.56\pm0.05$ & $1.14\pm0.01$ & $0.60\pm0.03$\\
1739 \AA& $5.62$ & $7.56\pm0.25$ & $2.65\pm0.25$ & $1.20\pm0.04$ & $0.76\pm0.12$\\
UVW2 & $6.04$ & $5.98\pm0.14$ & $2.12\pm0.09$ & $1.00\pm0.02$ & $2.10\pm0.05$\\
UVM2 & $7.27$ & $6.90\pm0.24$ & $2.32\pm0.23$ & $1.12\pm0.03$ & $2.86\pm0.10$\\
UVW1 & $7.59$ & $6.31\pm0.24$ & $2.13\pm0.17$ & $1.01\pm0.3$ & $3.62\pm0.08$\\
Swift U & $8.61$ & $6.00\pm0.25$ & $1.73\pm0.17$ & $0.95\pm0.03$ & $4.89\pm0.09$\\
Swift B  & $7.73$ & $4.08\pm0.06$ & $1.37\pm0.09$ & $0.86\pm0.02$ & $4.56\pm0.07$\\
Swift V  & $6.71$ & $4.45\pm0.14$ & $1.48\pm0.13$ & $0.65\pm0.01$ & $4.13\pm0.03$\\
$u$ & $7.93$ & $5.13\pm0.29$ & $1.45\pm0.18$ & $0.69\pm0.03$ & $5.21\pm0.08$\\
$B$ & $7.83$ & $4.32\pm0.21$ & $1.11\pm0.18$ & $0.62\pm0.01$ & $5.30\pm0.01$\\
$g$ & $10.38$ & $3.88\pm0.07$ & $1.22\pm0.09$ & $0.75\pm0.01$ & $7.38\pm0.02$\\
$V$ & $8.81$ & $3.94\pm0.26$ & $1.21\pm0.21$ & $0.61\pm0.03$ & $6.41\pm0.02$\\
$r$ & $11.12$ & $3.94\pm0.12$ & $1.20\pm0.09$ & $0.63\pm0.01$ & $8.60\pm0.02$\\
$i$ & $10.62$ & $3.54\pm0.11$ & $0.90\pm0.09$ & $0.55\pm0.01$ & $8.39\pm0.02$\\
$z$ & $12.11$ & $3.47\pm0.35$ & $0.70\pm0.20$ & $0.51\pm0.02$ & $10.02\pm0.05$\\
\enddata
\label{table:flxflx}
\tablecomments{Units for all columns except the filter bands are mJy. The last two columns are the additive shift $A_\nu$ (the constant component of the light curve) and the scaling value $S_\nu$ for the variable component. Bright and faint values are the maximum and minimum values of the light curve after subtraction of the constant component, while the mean flux values do not have the constant component removed.}
\end{deluxetable*}

\begin{figure*}[hbtp]
    \centering
    \includegraphics[width=\linewidth]{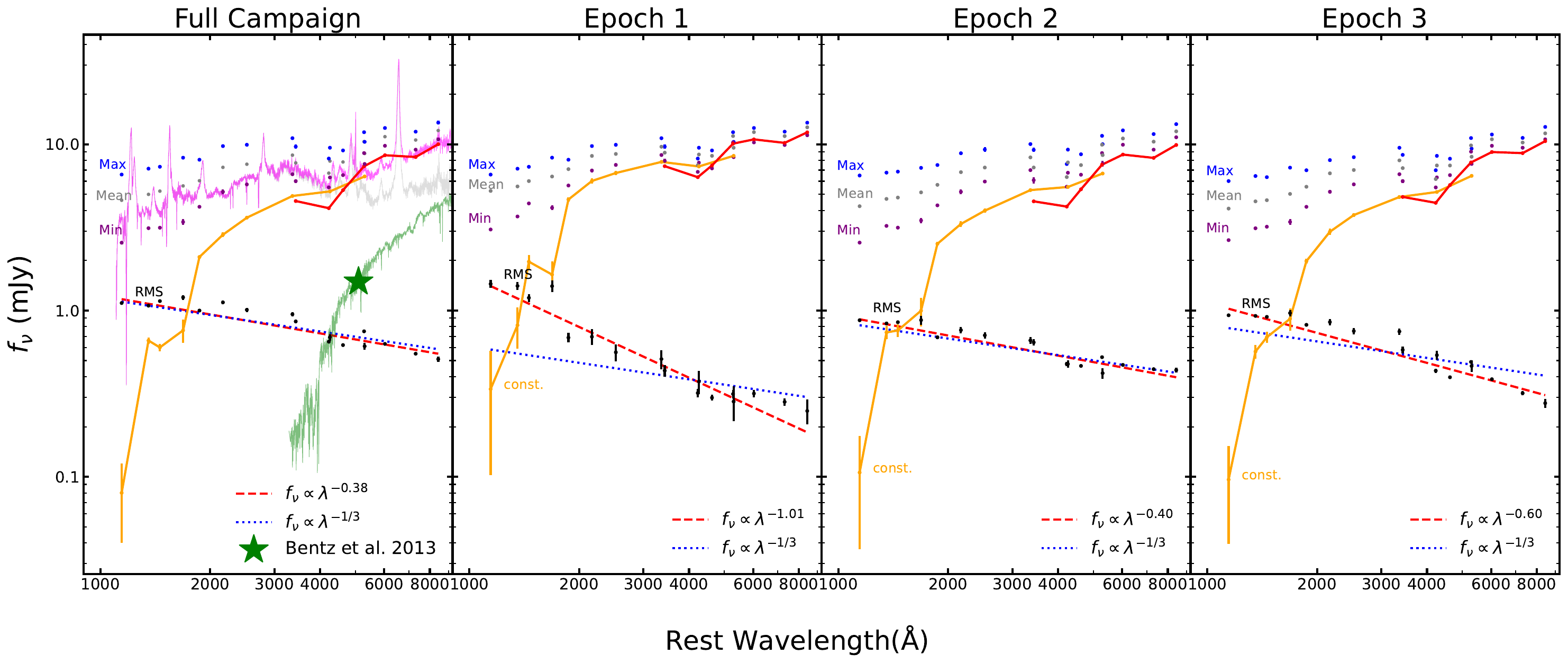}
    \caption{The broad-band spectral components of Mrk\,817 inferred from the flux-flux analysis for the full campaign (\emph{left panel}) and for the three epochs. The rms spectrum is fitted with an $f_\nu \propto \lambda^{-1/3}$ model (the blue-dotted line) as expected for an accretion disk, and with an $f_\nu \propto \lambda^\alpha$ model with exponent $\alpha$ as a free parameter (the red-dashed line). Maximum, mean, and minimum values for the total flux in each band are shown at the top of the plot. The solid orange line represents the constant component inferred for the space-based bands, and the solid red line represents the constant component for the ground-based bands.  In the left panel, the green star denotes the host galaxy flux at  5100 \AA\ in a $5\arcsec\times12\arcsec$ rectangular aperture, from \citet{Bentz2013}, and the green spectrum is a \citet{Bruzual_Charlot_SSP} 11 Gyr model normalized to match the host galaxy flux at 5100 \AA.  The STIS spectrum of Mrk 817 is shown in light gray (the same spectrum as shown in Figure \ref{fig:specfilter}, obtained with a 0\farcs2 slit width), and the sum of the STIS spectrum and the host galaxy template is shown in magenta. Comparison of the flux-flux results with the AGN+host spectral model demonstrates that the flux-flux constant component is predominantly AGN light across the full spectral range.}
    \label{fig:3ff}
\end{figure*}

Figure \ref{fig:3ff} shows the variable and constant component spectra derived for Mrk 817 (the left panel shows the spectra derived for the full STORM 2 campaign duration). Our results for the space-based bands reproduce the earlier measurements shown in Paper IV. For the region of wavelength overlap between the Swift and ground-based bands (the UBV bands), we find reasonable agreement between the rms flux values for the corresponding Swift and ground-based bands. Fitting a power law to the rms variability spectrum, we find a best-fit model of $f_\nu \propto \lambda^{-0.38}$, close to the characteristic $\lambda^{-1/3}$ spectrum of a standard accretion disk over the optical wavelength range.

\subsection{Individual epochs}
\label{sec:ff-epochs}
To check whether there are detectable variations in the variable and constant spectral components across the duration of the campaign, we applied the flux-flux analysis to the three individual epochs described in Section \ref{sec:epochlag}. The results are shown in Figure\,\ref{fig:3ff} for comparison with the full campaign measurements. The primary result is that the inferred variable and constant spectral components during Epoch\,1 differ substantially from those measured over the full campaign and for the second and third epochs. During Epoch\,1, the rms variability spectrum is best fit by a bluer $f_\nu \propto \lambda^{-1.01}$ power law, in contrast to the best-fitting power-law indices of $-0.38$ for the full campaign, and $-0.40$ and $-0.60$ for Epochs\,2 and 3. This appears to be a manifestation of the decoupling between the far-UV and near-UV variations that occurred during the early portion of the campaign. As illustrated in Figure 6 of \citet{cackett_storm2_IV}, the 1180 \AA\ and UVW2 light curves exhibit similar shapes over most of the STORM 2 campaign except during THJD $\approx9220-9320$, when the 1180 \AA\ light curve experienced a strong drop in flux followed by a recovery to its ``normal'' behavior tracking the UVW2 variations. This anomalous decline and subsequent recovery of the far-UV flux during Epoch\,1 increased the variability amplitude of the far-UV bands relative to the near-UV, and these changes can be associated with the steeper slope of the flux-flux rms component's spectrum during Epoch\,1. As discussed by \citet{cackett_storm2_IV}, this anomalous change in spectral shape could result from enhanced extinction related to the launch of the obscuring outflow in Epoch 1, or might be due to intrinsic fluctuations in the disk.

\begin{figure}
    \centering
    \includegraphics[width=\linewidth]{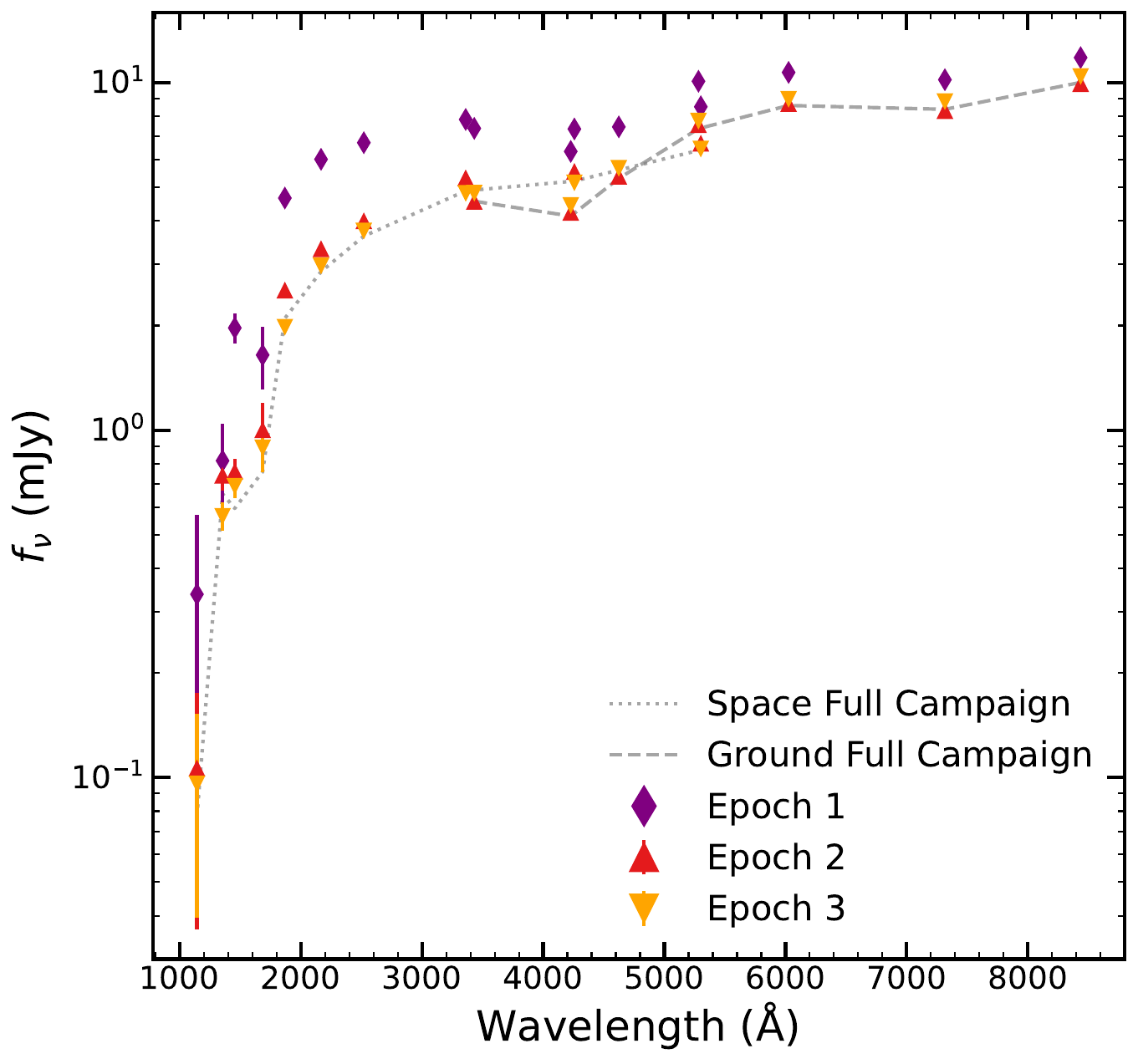}
    \caption{Comparison of the constant spectral components inferred for the three epochs (individual points) and for the full campaign duration (dashed and dot-dashed curves, representing the ground-based and space-based bands, respectively).  The Epoch 1 spectrum exhibits the largest difference from the full-campaign spectrum, particularly at UV wavelengths.}
    \label{fig:tripff}
\end{figure}

We also find that the inferred constant spectral component for Epoch\,1 deviates strongly from the constant component measured for the full campaign and for the later epochs. 
In earlier work using the flux-flux technique, it has sometimes been assumed that this so-called ``constant'' component primarily represents host-galaxy starlight, and the spectrum of this component has been referred to as the ``galaxy'' contribution to the overall spectrum \citep[e.g.,][]{fairall9juan, Weaver2022}. Our results demonstrate that this inferred constant component can actually change substantially as the AGN varies, implying that in Mrk 817 it is dominated by slowly varying contributions of AGN light rather than by starlight. To illustrate these changes, Figure \ref{fig:tripff} presents the constant component spectra for the three individual epochs and for the full campaign duration. While the Epoch 2 and 3 spectra are quite similar to the full campaign spectrum, the constant component spectrum for Epoch\,1 has a substantially higher flux across the spectrum, with the largest fractional change in the near-UV region. 

As an additional check, we re-measured the constant component spectrum for each epoch based on the value of $X_G$ obtained from the full-duration campaign, rather than using different $X_G$ values for each epoch. The results showed that the constant component spectrum derived for Epoch\,1 still differed significantly from those of Epochs\,2 and 3, similar to the results shown in Figure \ref{fig:tripff}, demonstrating that the changes in the constant component spectrum are not just caused by epoch-to-epoch variations in $X_G$ as derived from the 1180 \AA\ band.

Previously, \citet{cackett_storm2_IV} noted that the near-UV emission in the constant component from the flux-flux analysis of the full campaign duration was too strong to be accounted for by host galaxy light alone, and suggested that it might arise from diffuse continuum or \ion{Fe}{2} emission that varied more slowly than the $\sim$few-day reverberation timescale. This parallels the closely related conclusion from \citet{hagai_storm2_X} that the bluer-when-brighter trend seen in the HST spectroscopic data was intrinsic to the AGN and not merely the result of dilution by constant host galaxy light (although host dilution must play at least some role).

To follow up on these earlier findings, we compared the flux-flux constant component strength with the host galaxy flux measured by \citet{Bentz2013}, based on the HST image decompositions presented earlier by \citet{2009ApJ...697..160B}. In a $5\times12$ arcsec$^2$ rectangular aperture, corresponding to 76\% of the area of the $r=5\arcsec$ circular photometric aperture used for the STORM 2 Swift and ground-based photometry, they measured a host galaxy flux density of 1.49 mJy at 5100 \AA. This is a factor of 4.30 or 4.95 fainter than the Swift and ground-based $V$-band constant component fluxes determined from the flux-flux analysis, confirming that host galaxy starlight contributes only a small fraction of the flux-flux constant component spectrum. The host galaxy flux from \citet{Bentz2013} is shown as a green star in the left panel of  Figure \ref{fig:3ff}. We also show the spectrum of an 11 Gyr-old, Solar-metallicity stellar population from \citet{Bruzual_Charlot_SSP}, normalized to match the measured 5100 \AA\ galaxy flux, to illustrate a simple spectral model for the host galaxy bulge in comparison to the flux-flux constant component. Even if the actual host galaxy spectrum has a bluer spectral shape than this 11 Gyr model, there is no plausible host galaxy spectrum that could account for the majority of the flux-flux constant component given that it must be normalized at a value close to the directly measured host galaxy flux at 5100 \AA. \added{This result mirrors recent findings by \citet{Cai2024} that the constant component from the flux-flux method has a tendency to overestimate the host galaxy flux, although the discrepancy we obtain for Mrk 817 is larger than the factor of $\sim$1.5--2 discrepancies that \citet{Cai2024} found for other AGN.}

It is possible that the \citet{Bentz2013} measurement might somewhat underestimate the actual host galaxy flux, if Mrk 817 contained a nuclear star cluster so compact that it would be indistinguishable from the AGN point source in the GALFIT decomposition. However, no plausible nuclear star cluster could account for the large difference between the HST-derived value of the host galaxy flux and the value of the flux-flux constant component at the $V$ band. If the flux-flux constant component was dominated by starlight, the amount of additional host galaxy light that would be needed above the \citet{Bentz2013} value is roughly equivalent to the 5100 \AA\ flux of an entire $L_{\star}$ galaxy. Moreover, the putative nuclear star cluster would have to be contained within $r \lesssim 20$ pc to be hidden within the AGN PSF in the HST imaging. Any realistic flux contribution from an unresolved nuclear cluster could only have a minor impact on the normalization of the host galaxy model as shown in Figure \ref{fig:3ff}. 

Figure \ref{fig:3ff} also includes the HST STIS spectrum from 2022 January 2 (THJD 9581; the same spectrum shown in Figure \ref{fig:specfilter}), taken when Mrk 817 was close to its lowest flux during the STORM 2 campaign. We created a simple model for the combined AGN + host galaxy spectrum by adding the STIS spectrum (which is dominated by the AGN) and the normalized \citet{Bruzual_Charlot_SSP} model spectrum, illustrated in pink in the left panel of Figure \ref{fig:3ff}. At UV wavelengths the host galaxy contribution from the old stellar population is negligible. There may be some additional contribution to the host galaxy spectrum from a younger stellar population, but it is still likely to be very small in comparison with the AGN flux. The AGN+host model shows reasonably good agreement with the minimum-state spectrum obtained from the flux-flux analysis over most of the UV-optical spectral range.  Comparison of this AGN+host model spectrum with the flux-flux constant component further confirms that the constant component is dominated by AGN light at UV and blue wavelengths, with the host galaxy contribution increasing at red wavelengths to account for $\sim40\%$ of the constant component at the longest wavelengths. 

While it is clear that the linear flux-flux model can adequately fit the Mrk 817 light curve data over the luminosity range probed in this campaign, the physical interpretation of the constant component is ambiguous, and the only clear conclusion is that it is not dominated by the host galaxy. Our results highlight fundamental limitations in the flux-flux method: the relationship between light curve shapes in different bands can be more complex than the simple scaling described by Equation \ref{eqn:fluxflux}, since at any wavelength there are multiple contributions to the variable flux that can vary on widely different timescales and with different lags.  Additionally, Paper VI \citep{neustadt_storm2_VI} presented evidence for slowly propagating radial temperature fluctuations in the disk of Mrk 817, and the flux-flux model of Equation \ref{eqn:fluxflux} cannot account for the more complex spectral variability caused by such fluctuations. Finally, the method assumes that the variable spectral component has a constant color, so it cannot accommodate a variable spectrum that exhibits intrinsic ``bluer-when-brighter'' behavior such as has been inferred from quasar variability surveys \citep[e.g.,][]{VandenBerk2004,Sun2014}. As a result, the flux-flux method may not separate AGN and host galaxy light as cleanly as has been assumed in some prior work \added{\citep[see][for further discussion of this point]{Cai2024}}.   In future work on the flux-flux method, it would be worthwhile to explore whether the method can be extended to incorporate two (or more) time-variable components varying on different timescales, as a generalization of the model of Equation \ref{eqn:fluxflux}, in order to distinguish between rapidly and slowly varying spectral components in the data and separate them from the genuinely constant host-galaxy contribution.

\subsection{``Bowl'' Model}\label{sec:bowlmodel}

The Bowl model \citep{starkey_bowl,Prince2025} retains blackbody reprocessing but replaces the zero-thickness disk geometry with a concave power-law disk thickness profile following a relation $H\propto R^\beta$ between disk height and radius.
By tilting the outer disk reprocessing surface toward the central source of irradiation, the blackbody temperature is increased above the thin-disk profile $T\propto R^{-3/4}$. 
This shifts to shorter wavelengths the reprocessed light from the outer-disk annuli, giving larger lags at each wavelength.
The model parameters are constrained using MCMC methods to fit simultaneously the multi-band lag data and the AGN disk spectral data at the faintest and brightest levels observed during the campaign. A preliminary Bowl model fit to the HST and Swift data was presented by \citet{cackett_storm2_IV}, and here we describe a new fit incorporating the ground-based data to test whether this model can successfully match the observed spectral shape and lag spectrum.

\begin{figure*}[ht]
    \centering
    {\bf a)}
\includegraphics[trim={10mm 10mm 60mm 10mm}, clip, width=0.45\linewidth]{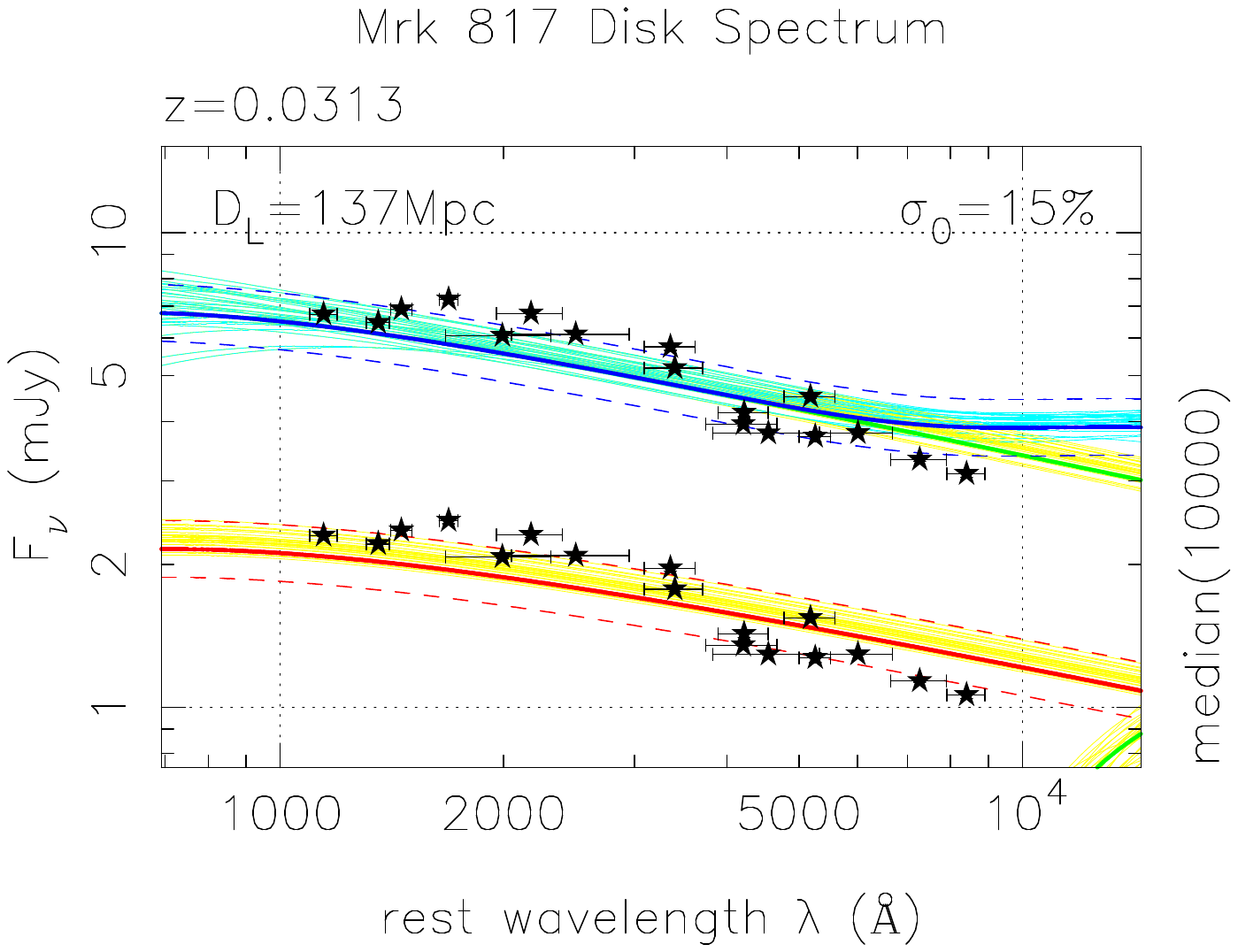}
    {\bf b)}
\includegraphics[trim={10mm 10mm 60mm 10mm}, clip,width=0.45\linewidth]{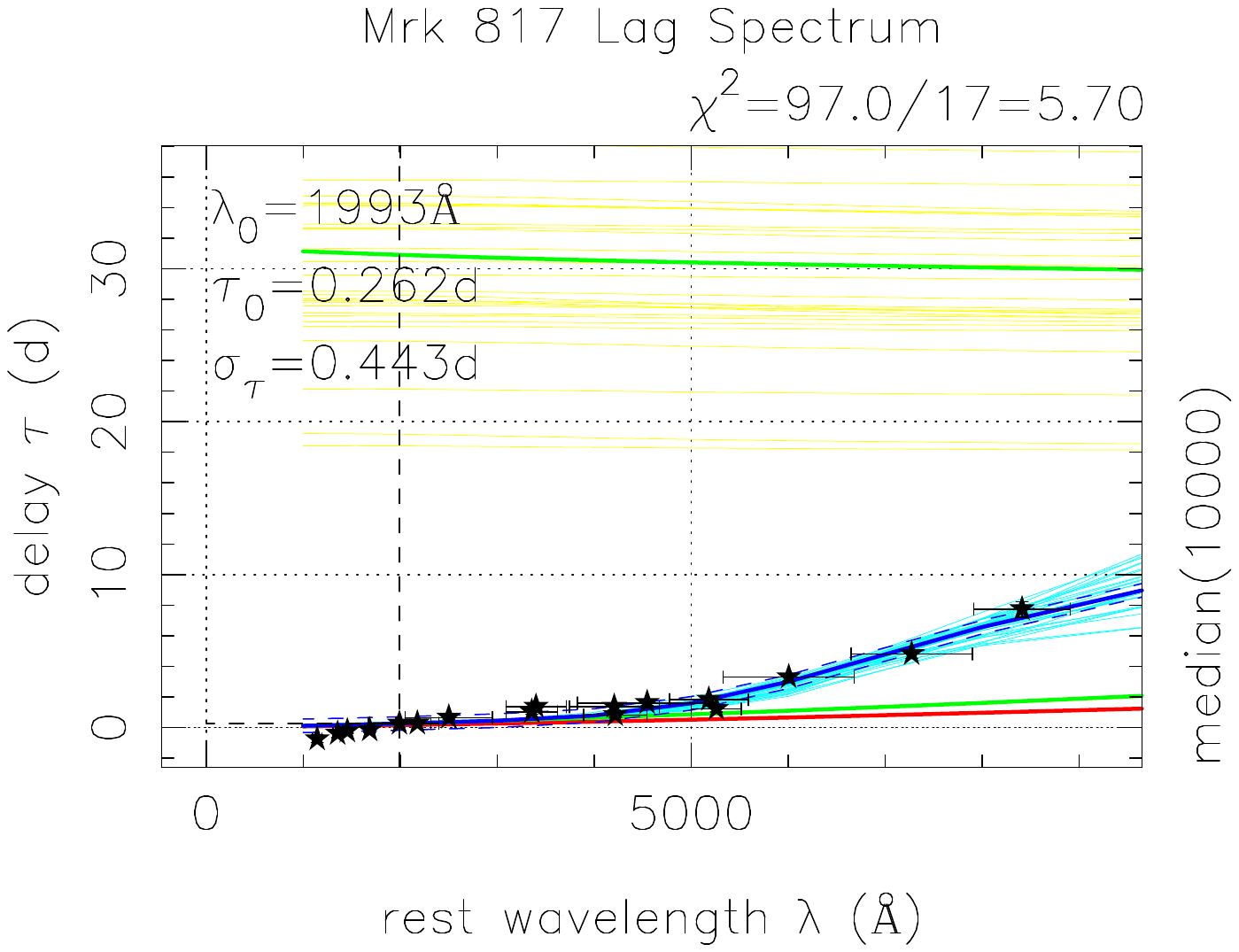}
\\
    {\bf c)}
\includegraphics[trim={10mm 10mm 60mm 10mm}, clip,width=0.45\linewidth]{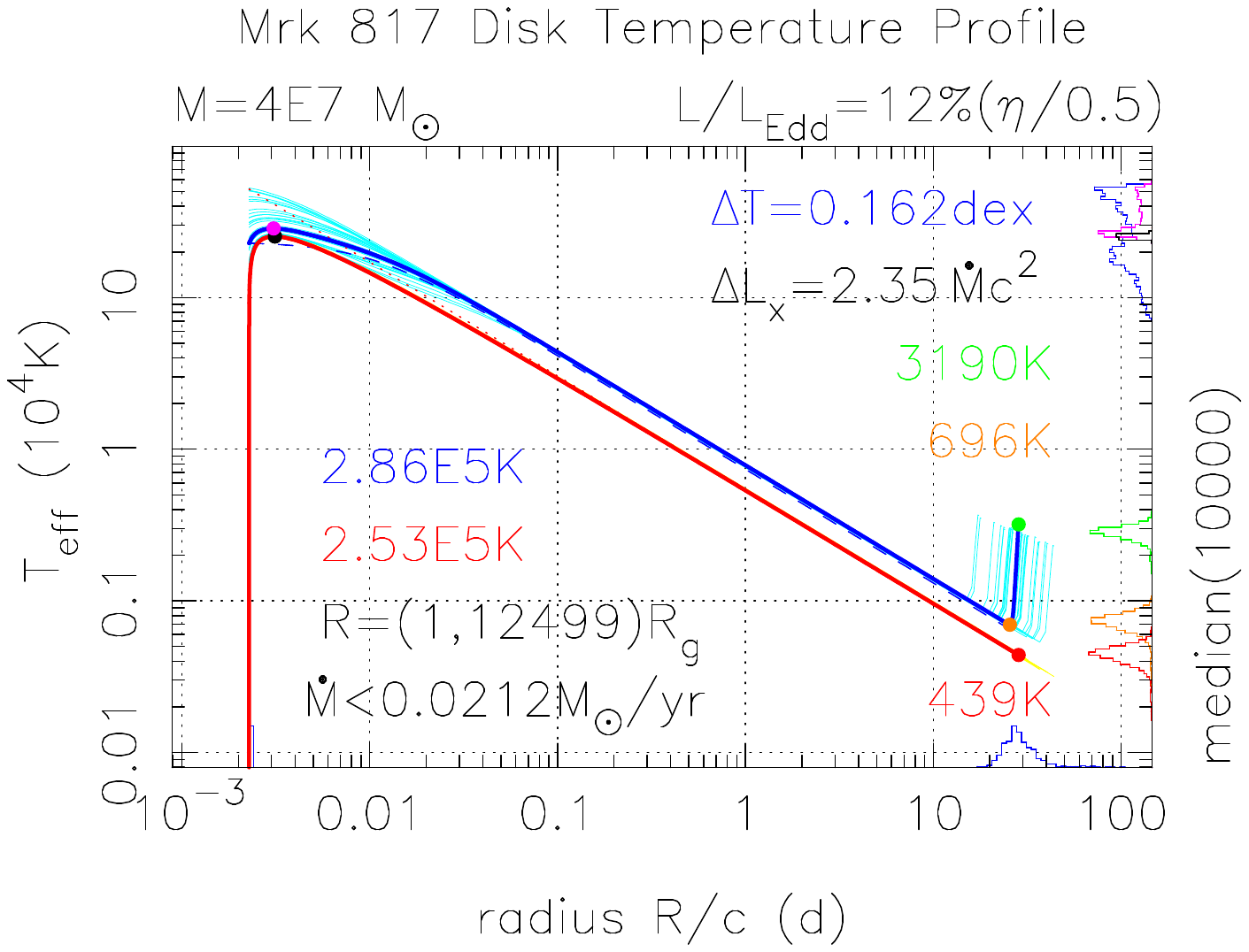}
    {\bf d)}
\includegraphics[trim={10mm 10mm 60mm 10mm}, clip,width=0.45\linewidth]{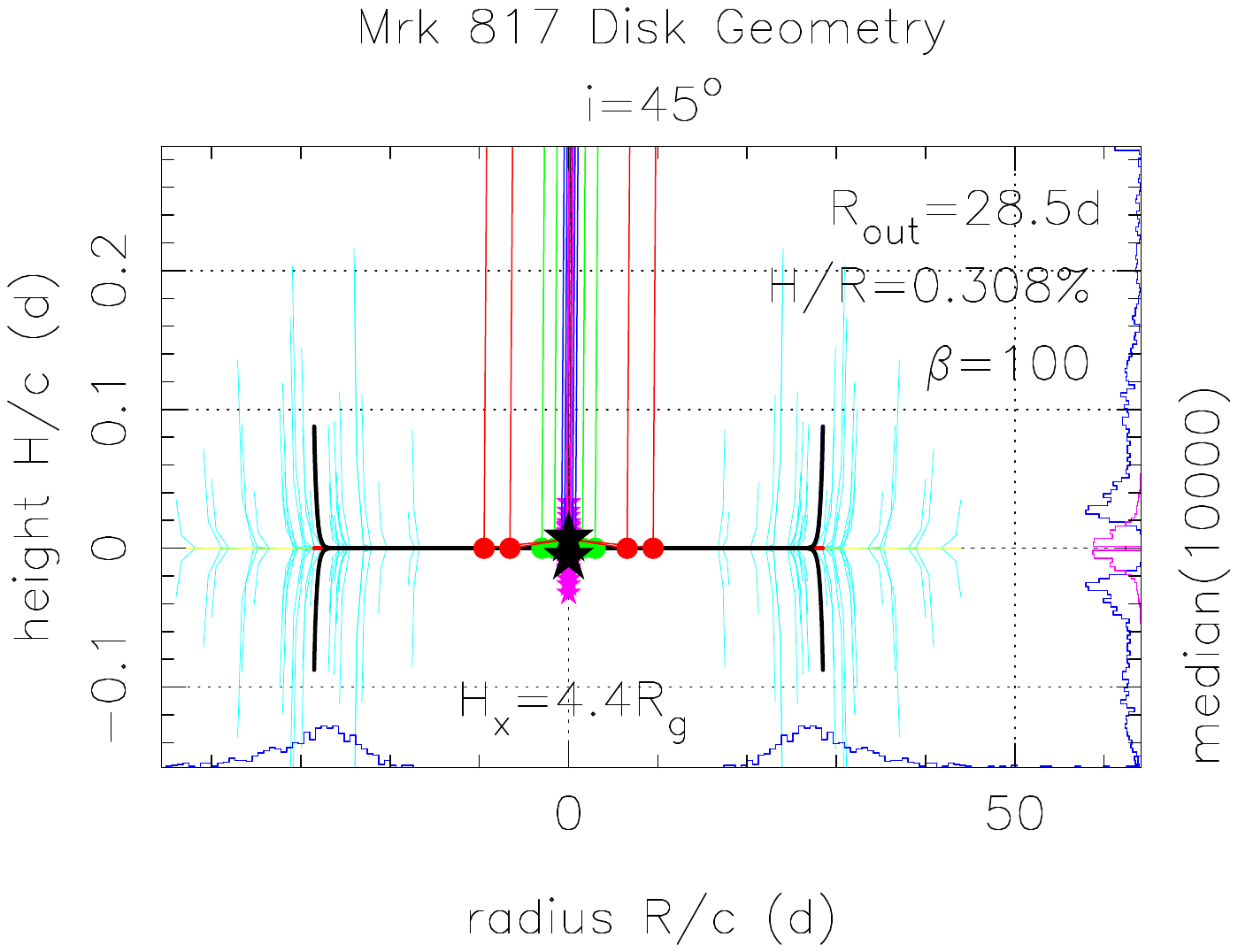}
    \caption{Results of fitting the Bowl model simultaneously to the  full-campaign faint and bright AGN \added{disk spectra (Panel\,a) and the PyROA lags (Panel\,b)}.
    The disk geometry \added{(Panel\,d)} has a steep outer rim,
    resulting in the temperature  profile \added{(Panel\,c)} falling as $T\propto R^{-3/4}$ and then rising on the outer rim.
    Red and blue curves correspond to the faint and bright state of the irradiated disk\added{, respectively}.
    Green curves in the top panels show the AGN broad-band spectrum and lag spectrum separately for the disk inside and outer edge outside the minimum temperature described in the model.
    A random selection of 30 MCMC samples (yellow and cyan) indicate uncertainties.
    \added{Parameter values given on the plot are medians of $10^4$ MCMC samples (see Section\,\ref{sec:bowlmodel} for details).}
    \added{In Panel\,c, the faint temperature profile (red) corresponds to the accretion rate upper limit $\dot{M}$ for the black hole mass $M$ and Eddington ratio $L/L_{\rm Edd}$ indicated on the plot.
    The bright temperature profile (blue) is higher by $\Delta T$, corresponding to the indicated $\Delta L_{\rm x}$.} 
    \added{Colored dots on the faint and bright temperature profiles mark the maximum and minimum temperatures, and the temperature at the rim, with similarly colored median values and histograms on the right edge indicating uncertainties based on the MCMC samples.
    In Panel\,d, colored lines trace light rays from the lamp to the disk and up to the observer at $i=45^\circ$, and histograms on the right edge indicate uncertainties in the lamp height $H_{\rm x}$ and the rim height $H(R_{\rm out})$, in pink and blue respectively.
    Blue histograms on the lower edges 
    in Panels\,c and d indicate uncertainty in the outer radius $R_{\rm out}$. }
    \label{fig:bowl}}
\end{figure*}

Figure\,\ref{fig:bowl} presents the results of fitting the Bowl model to the PyROA lags in Table\,\ref{tab:lags}, simultaneously with the faint and bright AGN disk spectral data in Table\,\ref{table:flxflx} from the flux-flux analysis shown in Figure\,\ref{fig:fluxflux}.
\added{We opt to use PyROA lags because the PyROA fit to multi-band light curves provides faint and bright AGN SEDs needed for the Bowl model fit. 
The CCF and JAVELIN methods give similar lags (Figure\,\ref{fig:lagwave}) but do not provide SED results.}

\added{Bowl model} parameters held fixed include 
the luminosity distance $D_{\rm L}=137$\,Mpc, black hole mass $M_\textrm{BH}=4\times10^7$\,M$_\odot$,
disk inclination $i=45^\circ$ as a fiducial value,
and inner disk radius $R_{\rm in}=R_g$ \added{($R_g=GM/c^2$ being the characteristic gravitational radius)}. 
The MCMC fit drives the disk thickness power-law index $\beta$ to large values, producing a flat disk with a steep outer rim. Fixing $\beta=100$ picks out a typical representative of these geometries.
Median parameters from $10^4$ MCMC samples are reported on the plot. 
The seven primary fit parameters
are 1) the disk rim radius $R_{\rm out}\sim 28$\, light days,
2) the rim height $H/R\sim 0.3\%$,
3) the central irradiating lamp height $H_x\sim 4.4\, R_g$, 
4) an upper limit on the accretion rate $\dot{M}\lesssim0.02$\,M$_\odot$\,yr$^{-1}$, 
5) the irradiation-driven temperature rise $\Delta T\approx0.16$\,dex, and rms uncertainties
6) $\sigma_0\approx15\%$ for the spectrum, and 7) $\sigma_\tau\approx0.44$\,d for the lags.
The noise model parameters $\sigma_0$ and $\sigma_\tau$, added in quadrature with the spectrum and delay measurement uncertainties, serve to quantify the fit residuals.
The MCMC samples indicate that
these parameters are approximately independent, apart from a tight correlation between $H_x$ and $H/R$.

The Bowl model fit reproduces the blue power-law slope of the AGN disk spectrum, the flux ratio between the faint and bright states, and the lags increasing with wavelength by about 8 days between 1000 and 9000\,\AA.
The lags in this blackbody reprocessing model arise from a wavelength-dependent mix of prompt ($\tau<1$\,d) response from the flat disk, which dominates in the UV, and  delayed response ($\tau\approx30$\,d) from the steep outer rim. 

The fit is not entirely satisfactory, however, in several respects. 
First, the spectrum data are slightly bluer than the $F_\nu\propto\nu^{1/3}$ spectrum of the irradiated thin-disk model (Figure \ref{fig:bowl}, panel a).
The model spectrum also turns up on the red end due to reprocessing on the inward face of the steep rim.
Finally, while in good
accord with the optical lag data,
the model lags are too flat across the UV, where HST lags are rising with wavelength.
The Bowl model fit is strained by this tension between fitting the red lags and avoiding a red upturn in the spectrum.

We note, however, that the Bowl model
attempts to fit the full lag spectrum entirely as the result of blackbody reprocessing from the disk surface, with no contributions from line and nebular continuum emission from the BLR or dust emission from the torus.
A more elaborate model incorporating one or both of these features could help to reduce this tension.
For example, hot dust at the inner edge of the torus may contribute to the lags in the reddest optical bands. 
A Bowl model with a second increase in thickness at that larger radius could be more successful.
The first step in disk thickness could then move in from 30 to 10 light days, where a jump in thickness is expected as dust in the disk atmosphere evaporates.
Here the wavelength dependence of \ion{H}{1} bound-free plus free-free opacity should modulate the height of the steep rim and produce corresponding jumps in the lag spectrum.
These elaborations (and others) are possible but beyond the scope of the current paper.

\section{Discussion} \label{sec:discussion}

\subsection{Summary of Measurements}

Earlier work in this series \citep{cackett_storm2_IV} presented reverberation lags and a flux-flux analysis for the HST and Swift continuum light curves of Mrk 817 measured during the STORM 2 campaign. In this paper, we add multiband ground-based photometric data to extend the wavelength coverage and carry out lag and flux-flux measurements over the three separate epochs during the campaign as defined by \citet{lewin_storm2_VII}, which had different levels of mean obscuration and luminosity.

Overall, we find that the optical light curves are highly correlated with the Swift UVW2 band, with $r_\mathrm{max} = 0.94-0.96$ for the $u$  through $i$ bands, and 0.88 for the $z$ band. Cross-correlation functions for each band relative to UVW2 exhibit asymmetries toward longer lags beyond the CCF peak, and the centroid lags $\tau_\mathrm{cen}$ are consistently greater than the peak lags $\tau_\mathrm{peak}$. These trends can be attributed to structure in the transfer functions extending toward lags beyond the peak lag, likely resulting from \added{diffuse continuum and emission lines from the BLR.}

For measurements over the full campaign duration, our PyCCF lag measurements result in a lag spectrum that is well fitted by the typical $\tau\propto\lambda^{4/3}$ power-law model (Equation \ref{eqn:powerlaw}), and we find evidence of an excess lag in the $u$ band, confirming the U-band excess seen earlier in the Swift data.  Such excess lags in the U  or $u$ bands have been seen in numerous other AGN to date \citep[e.g.,][]{Fausnaugh_2016,Cackett_2018_4593,Cackett2020} and are attributed to the contribution of Balmer continuum emission from the BLR \citep{lauther_goad_korista,koristagoad}. Measurements with JAVELIN and PyROA, in contrast, find somewhat shorter lags overall and little or no $u$-band excess \citep[also consistent with the findings of][]{cackett_storm2_IV}, but the $z$-band lags are found to be outliers falling well above the best-fitting $\lambda^{4/3}$ models (Figure \ref{fig:lagwave}). The normalization $\tau_0$ of the power-law fits to the lag spectra indicates a typical ``disk size discrepancy'' similar to results found  for other AGN in intensive monitoring programs \citep[see][for further discussion]{CACKETT2021102557}. Our derived values of $\tau_0$ exceed the basic disk reprocessing prediction by factors of 3.0--6.2 for the three lag measurement methods. The $\tau_0$ values found in this work are higher than those measured previously for the space-based STORM 2 data by \citet{cackett_storm2_IV}, since the longer lags of the ground-based bands favor larger values of $\tau_0$.

\begin{figure}[t]
    \centering
    \includegraphics[width=\linewidth]{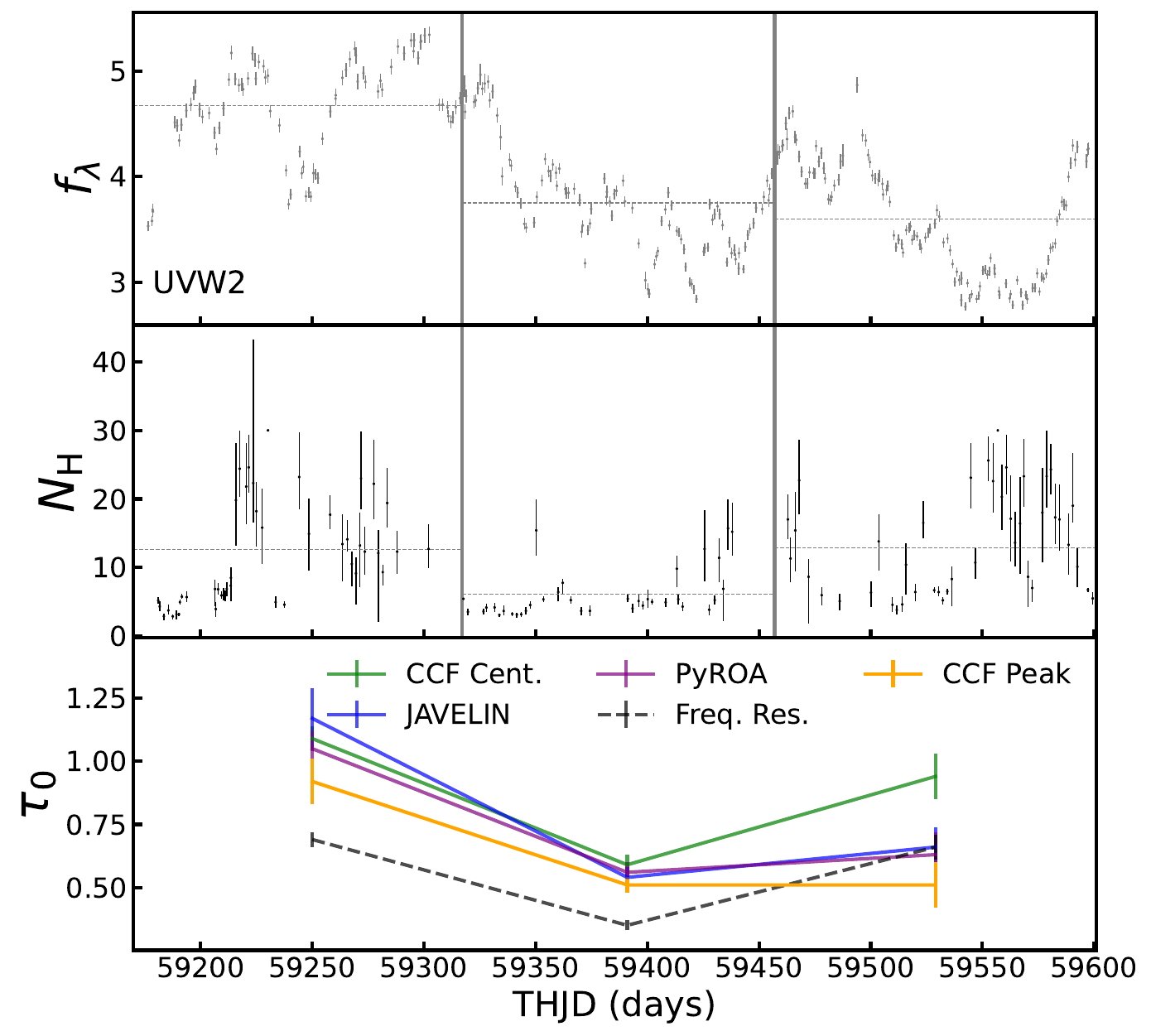}
    \caption{The Swift UVW2 light curve (top, in units of $10^{-14}$ erg cm$^{-2}$ s$^{-1}$ \AA$^{-1}$), X-ray absorbing column density $N_\mathrm{H}$ in units of $10^{-22}$ cm$^{-2}$ from \cite{Partington_2023_STORM2_III} (middle), and lag spectrum normalization factor $\tau_0$ (in days) for each of the three epochs during the campaign (bottom). In the upper and middle panels, gray horizontal lines show the mean values of UVW2 flux and $N_\mathrm{H}$ for each epoch. In the lower panel, in addition to the $\tau_0$ values measured in this work, we also show the $\tau_0$ values from the frequency-resolved lag analysis of Paper VII \citep{lewin_storm2_VII}, for the frequency range 0.014-0.048 day$^{-1}$.}
    \label{fig:uvw2_Nh_tau0}
\end{figure}

Recent work has found growing evidence for significant changes in AGN continuum lags across timescales of years from long-duration monitoring datasets \citep{Zhou2025,Su2025}.
The excellent data quality obtained during the STORM 2 program enables the detection of variations in continuum lags over much shorter timescales than have been probed in most other campaigns.
Figure \ref{fig:uvw2_Nh_tau0} illustrates the changes in $\tau_0$ across the three measurement epochs in comparison with the changes in the UVW2 flux and the X-ray absorbing column density $N_\mathrm{H}$ as measured from NICER data in Paper III \citep{Partington_2023_STORM2_III}. The lag measurements based on $\tau_\mathrm{cen}$ and from the frequency-resolved method of \citet{lewin_storm2_VII} show a drop from Epoch 1 to 2, followed by an increase in Epoch 3 to nearly the same lag as observed in Epoch 1,  mirroring the trend seen in $N_\mathrm{H}$ across the three epochs. In contrast, the $\tau_0$ values inferred from $\tau_\mathrm{peak}$, JAVELIN, and PyROA show the same drop from Epoch 1 to 2 but no significant rise again in Epoch 3, resembling more closely the trend seen in UV luminosity across the three epochs. These differing results likely stem from the relative sensitivity of the different lag measurement methods to the extended tail of the transfer function. If Epoch 3 is characterized by an increase in the DC emission fraction, corresponding to an enhancement of the long-timescale lags, this would be reflected in an increase in both $\tau_\mathrm{cen}$ and in the low-frequency component of the lag distribution as described by \citet{lewin_storm2_VII}, while $\tau_\mathrm{peak}$ might be largely unaffected by changes in the extended tail of the transfer function if it is primarily sensitive to prompt disk reverberation. Additionally, the different lag measurement methods may have different sensitivity to other, non-reverberating modes of variability such as disk temperature fluctuations \citep{neustadt_storm2_VI}.

Compiling high-quality continuum lag measurements for several AGN, \citet{Netzer2022} found a lag-luminosity relation of $\tau(5100\,\text{\AA})\propto L(5100\,\text{\AA})^{0.5}$ \citep[see also][]{Guo_2022}, mirroring the similar power-law trend found for broad emission-line lags \citep{Bentz2013}. Our epoch measurements allow us to examine whether the temporal changes in lag in Mrk 817 follow this scaling relation. We focus on the change between Epochs 1 and 2 since the AGN's mean luminosity and the continuum lags both undergo changes across these two periods. 

We use the $V$ band as a proxy for the rest-frame 5100 \AA\ continuum. After subtracting the host galaxy flux as measured by \citet{Bentz2013}, the mean $V$-band fluxes in Epochs 1 and 2 are $8.6\times10^{-15}$ and $7.4\times10^{-15}$ erg cm$^{-2}$ s$^{-1}$ \AA$^{-1}$, corresponding to a 14\% decrease from Epoch 1 to Epoch 2. If Mrk 817 followed the $\tau\propto L^{0.5}$ scaling as its luminosity varied, the expected change in lag would then be only $\sim7\%$. The flux decrease from the maximum $V$ flux of Epoch 1 to the minimum flux of Epoch 2 is 30\%.

To compare with the sample-derived lag-luminosity trend, we note that \citet{Netzer2022} derived the $\tau\propto L^{0.5}$ relation using adjusted 5100 \AA\ lags that were modified to use 1200 \AA\ (rather than UVW2) as the reference wavelength. For a fair comparison with the \citet{Netzer2022} relation, we add the $V$-band $\tau_\mathrm{cen}$ lag and the negative of the 1180 \AA\ $\tau_\mathrm{cen}$ lag as listed in Table \ref{epochlagstab} to derive an approximation of the lag between 5100 and 1200 \AA. This gives lags of $\tau(5100\,\,\text{\AA}) =$ 4.1 and 2.1 days for Epochs 1 and 2. This factor of $\sim2$ decrease in lag is far larger than what would be expected if Mrk 817 followed the $\tau\propto L^{0.5}$ scaling relation as it varied.  However, the Epoch 1 and 2 lags are both well within the very large scatter of the AGN sample in the $\tau$ vs.\ $L_\mathrm{5100}$ relation as shown in Figure 4 of \citet{Netzer2022}. Furthermore, $L_{5100}$ is a poor proxy for the ionizing UV luminosity in Mrk 817, which would have experienced a much greater drop from Epoch 1 to 2, and a more direct comparison of continuum lags with $L_\mathrm{UV}$ rather than $L_\mathrm{5100}$ for reverberation-mapped AGN might reveal lower scatter and clearer trends. Our results suggest that the dispersion in this relation could be caused in part by short-term fluctuations in continuum lags that may occur even at nearly constant 5100 \AA\ luminosity. Repeat measurements of continuum reverberation across multiple epochs in a larger sample of AGN will be needed to determine how common such fluctuations are, and would enable measurement of lags over epochs selected to correspond more closely to minima or maxima in luminosity which could provide more detailed understanding of the lag-luminosity relationship.

For the full campaign duration, the \emph{u}-band excess lag is detected only in the PyCCF $\tau_\mathrm{cen}$ measurements, while $\tau_\mathrm{peak}$, JAVELIN, and PyROA do not show this feature.  A puzzling aspect of the lags is that in the epoch-based $\tau_\mathrm{cen}$ measurements, the \emph{u}-band excess is clearly present only during Epoch\,3, an epoch characterized by relatively low UV and X-ray luminosity and elevated $N_\mathrm{H}$ \citep{Partington_2023_STORM2_III, cackett_storm2_IV}. However, Epoch 3 also shows the greatest discrepancy between $\tau_\mathrm{cen}$ and $\tau_\mathrm{peak}$, which as previously discussed may be an indication of enhanced DC emission contributing to the long tail of the transfer function. If so, this would provide a natural connection with the appearance of the \emph{u}-band excess lag. At present there are very few multi-epoch measurements of continuum lags in other AGN for comparison, but the behavior of Mrk 817 appears to be distinct from that seen in Mrk\,110, where the strength of the \emph{u}-band excess lag across multiple monitoring epochs was seen to scale with X-ray luminosity \citep{Vincentelli2022}.

While these results extend the previous results of Paper IV, one difference is that \citet{cackett_storm2_IV} measured the HST/Swift continuum lags both before and after applying a detrending procedure to the light curves to remove low-frequency structure. In reverberation mapping, a low-order (linear or sometimes quadratic) detrending is often used to remove an overall increasing or decreasing trend in flux, so that cross-correlation analysis can better isolate the lags due to short-timescale variations \citep{1999PASP..111.1347W}. 
In Mrk\,817, the temporary divergence between the far-UV and near-UV light curve shapes during Epoch 1 is problematic for lag measurement methods such as JAVELIN and PyROA that fit the light curves under the assumption that the responding bands can be modeled as time-shifted and broadened versions of the driving continuum band. To obtain adequate fits to the light curves with these methods, \citet{cackett_storm2_IV} detrended the data by smoothing the light curves with a Gaussian kernel of width $\sigma=20$ days and then subtracting the smoothed model from the data. This procedure amounts to a much higher-order correction than the linear detrending typically used in reverberation mapping.  After this detrending, the lags for the Swift UBV bands and the derived values of $\tau_0$ were shorter for all three lag measurement methods and close to the predictions for standard disk reprocessing. These results suggested that the longer-timescale ($\gtrsim20$ d) light curve variations produce contributions to the lags arising from nebular continuum and line emission in the BLR, while the shorter-timescale variations remaining in the detrended light curves are associated with lags arising on smaller spatial scales in the accretion disk.

Combining the HST, Swift, and ground-based light curves, \citet{lewin_storm2_VII} carried out an analysis in which the lags were measured as a function of temporal frequency of variations. They were able to measure the lags independently across six frequency bins covering a range of $7\times10^{-3}$ to 1.0 day$^{-1}$, finding a strong trend of decreasing lag as a function of frequency. Their results further support the conclusion that the lags corresponding to low-frequency variations are associated with reprocessing on the spatial scale of the BLR, while the lags occurring in response to high-frequency variations ($\gtrsim0.05$ d$^{-1}$) are associated with reprocessing at the scale of the disk. Since the work of \citet{lewin_storm2_VII} provides  a much more detailed view of the frequency-resolved lag behavior in the STORM\,2 campaign in comparison with the detrending method of \citet{cackett_storm2_IV}, we do not include detrended measurements as part of this work.

\subsection{Spectral Variations and the Impact of Disk Winds}

Based on the variations in mean obscuring column density $\bar{N}_\mathrm{H}$ across the three epochs, \citet{lewin_storm2_VII} proposed a scenario to explain the lag variations that connects changes in lag to time-varying obscuration between the BLR and the ionizing continuum source. In this model, the optical light curves contain two components: a disk reprocessing component producing short lags, and a diffuse continuum component from the BLR giving longer lags. Periods of low $N_\mathrm{H}$ (i.e., Epoch 2 of the STORM 2 campaign) correspond to times when an episodic, clumpy disk wind has just launched from the disk surface and has not yet intersected the observer's line of sight to the X-ray emitting region. The clumpy wind partially shields the BLR from the ionizing continuum, reducing the strength of the BLR continuum contribution so that the overall continuum lags are similar to those of the disk reprocessing component alone.  As the clumpy wind rises further above the disk surface, it intercepts our line of sight toward the AGN central engine, resulting in elevated $N_\mathrm{H}$, as seen during Epochs 1 and 3. During this phase the larger scale height of the clumpy wind lifts the shielding of the BLR from the ionizing continuum, and the fully illuminated BLR then produces a strong diffuse continuum signal resulting in longer lags overall. This model is broadly similar to the scenario proposed by \citet{homayouni_storm2_V} to explain temporal changes in the \ion{C}{4} emission-line lag over the duration of the campaign. 

Recently, \citet{Netzer2025} presented a different approach to explain the time-variable lags seen in the STORM 2 campaign and their relationship to the observed changes in $N_\mathrm{H}$. \citet{hagai_storm2_X} and N25 used HST spectra to demonstrate that the spectral shape of Mrk 817 changed during the STORM 2 campaign, exhibiting ``bluer-when-brighter'' behavior.  N25 proposed that these spectral changes result from disk winds episodically depleting gas from the accretion disk and changing its emitted spectrum, as discussed earlier by \citet{Slone2012}. Based on the X-ray observations presented by \citet{Partington_2023_STORM2_III} and \citet{zaidouni_storm2_IX}, N25 estimated the mass outflow rate of the disk wind in Mrk\,817 to be $\sim$0.1 M$_\odot$ yr$^{-1}$, a significant fraction of the accretion rate through the outer accretion disk.  The lowered accretion rate at small radii ($\lesssim100 R_g$) then decreases the inner disk temperature, softening the disk's UV spectral shape and reducing the ionizing luminosity by as much as a factor of $\sim$4 for the adopted disk and wind parameters. In the aftermath of a wind-launching episode, the ionizing photon flux incident on the BLR would then be strongly diminished, lowering the intensity of the diffuse continuum emission and reducing the observed continuum lags to values closer to the expected contribution from disk reprocessing. This mechanism operates without the need to invoke shadowing of the BLR by wind material to explain the decrease in lags from Epoch\,1 to Epoch\,2.

\begin{figure*}[hbtp]
    \centering
    \includegraphics[width=\linewidth]{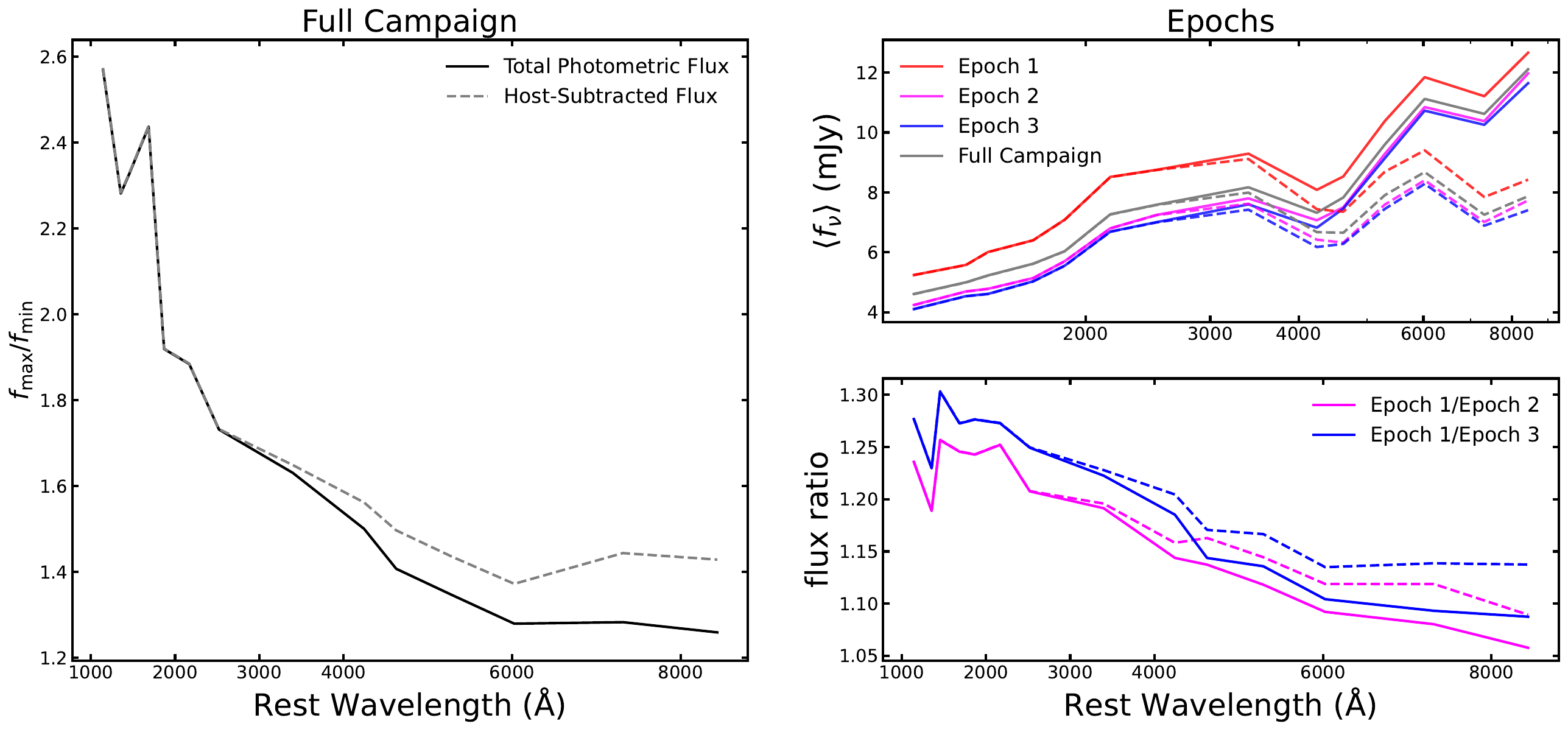}
    \caption{\emph{Left panel:} Ratio of maximum-state to minimum-state spectra. \emph{Upper right panel:} Mean spectra from the flux-flux analysis for the three epochs and for the full campaign. \emph{Lower right panel:} The ratio of the Epoch 1 spectrum to the spectra of Epochs 2 and 3. In each panel, solid and dashed curves represent the data before and after subtraction of the host-galaxy model, respectively. The overall slopes of the flux-ratio curves demonstrate that the mean spectrum has the bluest shape during Epoch 1, corresponding to the epoch of highest luminosity. In all panels we show an average of the Swift U and ground-based $u$ data points in each curve rather than plotting the ground and space-based points separately, and the same is done for the B and V bands. }
    \label{fig:ff_epoch_mean_comp}
\end{figure*}

Our flux-flux analysis provides another way to examine the broad-band spectral changes in Mrk 817 during the course of the campaign and their connection with the continuum lags. Figure \ref{fig:ff_epoch_mean_comp} (left panel) shows the ratio of the maximum-state to minimum-state spectra obtained from the flux-flux analysis, illustrating the full amplitude of the bluer-when-brighter spectral slope change over the duration of the entire campaign. Figure \ref{fig:ff_epoch_mean_comp} also shows the mean spectral shape for each epoch, and the ratio of the Epoch 1 mean spectrum to the mean spectra during Epochs 2 and 3 (both before and after subtraction of the host galaxy model spectrum).
These spectral ratios rise toward the UV, further confirming that the epoch-averaged broad-band spectra do show the bluer-when-brighter behavior, and that this trend is intrinsic to the AGN and not just the result of host galaxy contamination at longer wavelengths.

While our results confirm the occurrence of luminosity-dependent changes in Mrk 817's UV/optical spectral shape, implying a substantial drop in ionizing luminosity after the end of Epoch 1, the underlying causal connections between the wind-launching events and changes in obscuration, the AGN's spectral variations, and the changes in continuum lag across the duration of the campaign are difficult to ascertain.  Both the obscuration-based model of \citet{lewin_storm2_VII} and the wind-depleted disk model of \citet{Netzer2025} provide frameworks that can plausibly account for the decrease in continuum lags between Epoch 1 and Epoch 2.  The interpretation of the Epoch 3 observations also remains ambiguous. We observe an increase in $\tau_\mathrm{cen}$ from Epoch 2 to 3 despite a slight drop in UV luminosity, and this increase in lag would not be expected under in the \citet{Netzer2025} scenario. However, the other measures of lag ($\tau_\mathrm{peak}$, JAVELIN, and PyROA, remain nearly constant from Epoch 2 to Epoch 3, appearing to track the roughly constant luminosity. On the other hand, the obscuration increases in Epoch 3 to a mean level similar to Epoch 1, and the pattern of changes in $\tau_\mathrm{cen}$ (but not $\tau_\mathrm{peak}$, JAVELIN, or PyROA) appears similar to the variation in $\bar{N}_\mathrm{H}$ across the three epochs.

N25 identified the elevated $N_\mathrm{H}$ during Epoch 1 with the onset of the wind that subsequently caused the softening of the spectral shape and the reduction in lags observed during Epoch 2. Our measurements show that this change in spectral shape persisted through Epoch 3, during which the next period of elevated  obscuration occurred. If the reappearance of elevated $N_\mathrm{H}$ during Epoch 3 corresponded to a new wind-launching event similar to the one that occurred during Epoch 1,  then we might predict a further luminosity decrease and softening of the spectra to occur at a later time, after the end of the STORM 2 monitoring period (during the extended campaign). The extended campaign data show, however, that the AGN luminosity actually increased after the STORM 2 monitoring period ended (Edelson et al., in preparation), so the possible connection between outflow events and subsequent spectrum changes is not entirely clear. A longer monitoring campaign could provide firmer tests for a causal link by examining whether wind-launching events are generally followed by spectral changes in the manner proposed by N25. Fortunately, the extended campaign dataset will more than double the total monitoring duration for Mrk\,817, and will aid in searching for connections between luminosity, spectrum shape, obscuration, and reverberation lags over a longer time baseline to better understand the mechanisms underlying these temporal changes in continuum lags.

The extended campaign dataset will also make it possible to search for further episodes of anomalous behavior such as the period around THJD $\approx9220-9320$, when the far-UV flux was temporarily depressed in comparison with the shape of the near-UV and optical light curves. This decorrelation event coincided roughly with the period of elevated X-ray absorption during Epoch 1. This behavior could result from enhanced UV extinction due to a dusty outflow \citep[e.g.,][]{Czerny2017} temporarily obscuring a portion of the accretion disk. Identifying additional events through multiwavelength monitoring will aid in understanding the mechanism underlying such changes in the UV/optical SED.

Future work to model the transfer function for these continuum light curves, using techniques such as MEMEcho \citep[e.g.,][]{horne_storm1_IX} can help to clarify the origin of the continuum lags. The exceptional quality of the STORM\,2 dataset, and the additional temporal coverage provided by the extended campaign, will make it possible to search for time-dependent and wavelength-dependent structure in the continuum transfer function corresponding to the expected contributions of disk, BLR, and dust torus emission, giving a more comprehensive view of the reprocessing geometry in Mrk\,817.

\section{Conclusions} \label{sec:conclusion}
We have presented the measurements and results for the 1.4-year STORM\,2 ground-based photometric campaign of Mrk\,817. The primary data products of this work are the  intercalibrated light curves in the \emph{BV} and \emph{ugriz} bands incorporating data from 12 telescopes. In combination with light curves measured from HST and Swift observations, this is among the best-quality high-cadence continuum monitoring datasets obtained to date for any AGN.  We have measured time delays using PyCCF, JAVELIN and PyROA relative to the Swift UVW2 (1928 \AA) light curve. Our primary conclusions are as follows.

\begin{itemize}

    \item The lags increase with wavelength up to a maximum of $\sim8$ days for the \emph{z} band (Figure \ref{fig:epoch_lag_wave}). The lag spectrum exhibits a typical ``disk size discrepancy'' in which the lags are $\sim3-6$ times longer than basic disk reprocessing predictions, similar to results for several other AGN in earlier literature.  The $\tau_\mathrm{cen}$ measurements for the full campaign duration show an excess lag in the U and \emph{u} bands, likely due to Balmer continuum variability, but this feature is weak or absent in the $\tau_\mathrm{peak}$, JAVELIN, and PyROA measurements, and in the $\tau_\mathrm{cen}$ epoch-based measurements it only appears clearly during the third epoch.

\item   Dividing the campaign into three epochs following the earlier work of \citet{lewin_storm2_VII}, we find a large drop in reverberation lags from Epoch 1 to Epoch 2, corresponding to a factor of $\sim$2 decrease in the normalization $\tau_0$ of the lag spectrum. This change in lag accompanies a decrease in both UV luminosity and mean obscuration ($\bar{N}_\mathrm{H}$) from Epoch 1 to Epoch 2. The behavior during Epoch 3 appears to be more complex: we find an increase in lags relative to Epoch 2 based on the $\tau_\mathrm{cen}$ measurements, but no increase is seen with $\tau_\mathrm{peak}$, JAVELIN, or PyROA. The different outcomes from different lag measurement methods may indicate a change in shape of the transfer function, resulting from an increase in long-lag contributions from the DC component relative to the disk reverberation lags. The $u$-band excess lag is strongest during Epoch 3, further supporting this interpretation.

\item We carried out a flux-flux analysis (Figure \ref{fig:fluxflux}), finding changes in the derived shape of the variable and constant-component spectra in different epochs of the campaign (Figure \ref{fig:3ff}). The constant component derived from the flux-flux technique is typically presumed to represent the contribution of host galaxy starlight to the spectrum, but we show that its flux changes across the three measurement epochs, and that the host galaxy accounts for only a small fraction of the flux in this component. These results show that the inferred constant spectral component must be dominated by constant or slowly varying contributions of AGN light rather than host-galaxy starlight. We also show that the AGN spectrum exhibits bluer-when-brighter behavior, confirming earlier findings based on HST spectroscopic observations obtained during the STORM\,2 campaign \citep{hagai_storm2_X, Netzer2025}.  While the host galaxy may contribute up to $\sim40\%$ of the flux in our $r=5\arcsec$ photometric aperture at the longest observed wavelengths, the overall bluer-when-brighter trend across the full observed spectral range cannot be attributed to host galaxy dilution, and we conclude that the luminosity-dependent changes in spectral shape are primarily intrinsic to the AGN.

    \item We fit the data with the bowl-shaped disk reprocessing model of \citet{starkey_bowl}. While this model can roughly reproduce the shape of the lag spectrum and the variable spectrum shape inferred from the flux-flux analysis, the model strains to fit the long-wavelength spectrum in detail, and the fit requires a rather extreme and sudden change in the thickness of the outer disk rather than a smooth variation of disk thickness with radius. These issues can be further examined in future work by combining the bowl reprocessing model with additional contributions of variable continuum emission from the BLR and dusty torus. 

\item   The observed changes in continuum lag during the campaign appear to track a decrease in luminosity and softening of the spectrum between Epoch 1 and 2. These changes are compatible with the model presented by \citet{Netzer2025}, in which the onset of a disk wind lowers the accretion rate through the inner region of the accretion disk and reduces the disk's ionizing photon luminosity. The luminosity of the nebular continuum and broad emission lines drop in response to this change in ionizing luminosity, reducing the broad-band lags overall. However, during Epoch 3 the AGN spectrum remains in a fainter and softer state similar to that of Epoch 2, and the observed increase in $\tau_\mathrm{cen}$ lag during Epoch 3 does not match the expectations of the \cite{Netzer2025} model. Thus, the available evidence does not definitively show that the disk wind observed during Epoch 1 is responsible for the later changes in the spectrum and the continuum lags. A more complete understanding of the impacts of changing AGN luminosity, spectral shape, and obscuration on the continuum lags will require more such wind-launching events and large-amplitude flux variations to be detected in future long-duration monitoring of Mrk\,817 and/or other AGN.

\end{itemize}

Future work on continuum reverberation in Mrk\,817 will make use of data from the extended campaign. That program obtained Swift and ground-based monitoring for an additional two years after the end of the STORM\,2 HST observing program, supported by additional HST COS UV spectroscopy and X-ray observations with NICER. The combined 3.2-year dataset will be unprecedented in terms of monitoring duration and cadence. This will enable more detailed investigations of topics studied in the STORM\,2 campaign including frequency-resolved lag behavior \citep{lewin_storm2_VII}, disk temperature fluctuations \citep{neustadt_storm2_VI}, and temporal variations in continuum lag and their relationship to changes in AGN luminosity, spectral shape, and disk winds.

\begin{acknowledgments}
We thank the referee for the helpful suggestions that improved this work.

Support for HST program GO-16196 was provided by NASA through a grant from the Space Telescope Science Institute, which is operated by the Association of Universities for Research in Astronomy, Inc., under NASA contract NAS5-26555. 

This work makes use of observations from the Las Cumbres Observatory global telescope network. This paper is based in part on observations made with the MuSCAT3 instrument, developed by Astrobiology Center and under financial supports by JSPS KAKENHI (JP18H05439) and JST PRESTO (JPMJPR1775), at Faulkes Telescope North on Maui, HI, operated by the Las Cumbres Observatory. 

This work is also based on observations collected at Schmidt 67/92 telescope (Asiago Mount Ekar, Italy) INAF - Osservatorio Astronomico di Padova.

The Liverpool Telescope is operated on the island of La Palma by Liverpool John Moores University in the Spanish Observatorio del Roque de los Muchachos of the Instituto de Astrofisica de Canarias with financial support from the UK Science and Technology Facilities Council.

This work makes use of observations collected at the Centro Astron\'omico Hispano en Andaluc\'ia (CAHA) at Calar Alto, operated jointly by the Andalusian Unviversities and the Instituto de Astrofisica de Andalucia (CSIC). Funding for the Lijiang 2.4 m telescope has bene provided by the Chinese Academy of Sciences (CAS) and the People's Government of Yunnan Province.

Research at UC Irvine was supported in part by NSF grant AST-1907290.
D.I., A.B.K, and L.\v C.P. acknowledge funding provided by the University of Belgrade—Faculty of Mathematics (contract 451-03-136/2025-03/200104), Astronomical Observatory Belgrade (contract 451-03-136/2025-03/200002), through grants by the Ministry of Education, Science, and Technological Development of the Republic of Serbia. M.D. thanks NASA for the support provided by grants JWST-AR-06419, JWST-AR-06428, JWST GO5018, and JWST GO5354 from the Space Telescope Science Institute (STScI). CSK is supported by NSF grants AST-2307385 and AST-2407206. 
YRL acknowledges financial support from the NSFC through grant No. 12273041 and from the Youth Innovation Promotion Association CAS. H.L. acknowledges a Daphne Jackson Fellowship sponsored by the Science and Technology Facilities Council (STFC), UK, and financial support from STFC grants ST/P000541/1 and ST/T000244/1. M.V. gratefully acknowledges financial support from the Independent Research Fund Denmark via grant number DFF 3103-00146 and from the Carlsberg Foundation (grant CF23-0417).

\facility{Swift, LCOGT, FTN, Liverpool:2m, Zowada, YAO:2.4m, Wise Observatory, Asiago:Schmidt, BYU:0.9m, CAO:Schmidt}

\software{Astropy \citep{astropy:2022},
          Matplotlib \citep{Hunter:2007}, lmfit \citep{matt_newville_2023_8145703}, PyCCF \citep{2018ascl.soft05032S}, PyCALI \citep{Li_2014}, JAVELIN \citep{2010ascl.soft10007Z}, PyROA \citep{pyroa}, astrometry.net \citep{Lang_2010} }
          
\end{acknowledgments}

\clearpage
\bibliography{jwm.bib}{}
\bibliographystyle{aasjournal}

\end{document}